\documentclass[a4paper,11pt]{article}
\usepackage[utf8]{inputenc}

\usepackage{framed}
\usepackage{mdframed}

\usepackage{fullpage}
\usepackage[T1]{fontenc}

\usepackage{epsf}
\usepackage{graphicx}
\usepackage{verbatim}
\usepackage{hyperref}
\usepackage{color}	
\usepackage{bbold}

\usepackage[british]{babel}
\usepackage{verbatim}
\usepackage[T1]{fontenc}
\usepackage{lmodern}
\usepackage{lipsum}
\usepackage{booktabs}
\usepackage{caption}
\usepackage{cite}
\usepackage{soul,color}
\usepackage[toc,page]{appendix}

\usepackage{ifpdf}

\usepackage{amsthm, amssymb}
\usepackage[tbtags]{amsmath}
\usepackage{bm}
\usepackage{mathtools}
\usepackage{amstext}
\usepackage{braket}
\usepackage{multirow}

\usepackage[normalem]{ulem}
\usepackage{xcolor}

\newcommand\redsout{\bgroup\markoverwith{\textcolor{red}{\rule[0.5ex]{2pt}{0.4pt}}}\ULon}

\begin{document}
\vspace{5mm}
\vspace{0.5cm}

\def\be{\begin{eqnarray}}
\def\ee{\end{eqnarray}}

\def\ba{\begin{aligned}}
\def\ea{\end{aligned}}

\def\ls{\left[}
\def\rs{\right]}
\def\lc{\left\{}
\def\rc{\right\}}

\def\p{\partial}

\def\S{\Sigma}

\def\s{\sigma}

\def\O{\Omega}

\def\a{\alpha}
\def\b{\beta}
\def\g{\gamma}

\def\ad{{\dot \alpha}}
\def\bd{{\dot \beta}}
\def\gd{{\dot \gamma}}
\newcommand{\ft}[2]{{\textstyle\frac{#1}{#2}}}
\def\ib{{\overline \imath}}
\def\jb{{\overline \jmath}}
\def\Re{\mathop{\rm Re}\nolimits}
\def\Im{\mathop{\rm Im}\nolimits}
\def\trace{\mathop{\rm Tr}\nolimits}
\def\rmi{{ i}}

\def\N{\mathcal{N}}

\newcommand{\SU}{\mathop{\rm SU}}
\newcommand{\SO}{\mathop{\rm SO}}
\newcommand{\U}{\mathop{\rm {}U}}
\newcommand{\USp}{\mathop{\rm {}USp}}
\newcommand{\OSp}{\mathop{\rm {}OSp}}
\newcommand{\Symp}{\mathop{\rm {}Sp}}
\newcommand{\Sl}{\mathop{\rm {}S}\ell }
\newcommand{\Gl}{\mathop{\rm {}G}\ell }
\newcommand{\Spin}{\mathop{\rm {}Spin}}

\def\hc{c.c.}

\numberwithin{equation}{section}

\allowdisplaybreaks

\allowbreak



\begin{titlepage}
	\thispagestyle{empty}
	\begin{flushright}

	\end{flushright}
\vspace{35pt}

	\begin{center}
	    {\Large{ 
	Three-dimensional flux vacua from IIB on co-calibrated G2 orientifolds 
	     }}

		\vspace{50pt}

		{\large Maxim~Emelin$^{a,b}$, \ Fotis~Farakos$^{a,b}$, \ George~Tringas$^{c}$}

		\vspace{25pt}

		{
			 $^{a}${\it Dipartimento di Fisica e Astronomia ``Galileo Galilei''\\
			Universit\`a di Padova, Via Marzolo 8, 35131 Padova, Italy }

		\vspace{15pt}

			$^{b}${\it   INFN, Sezione di Padova \\
    		Via Marzolo 8, 35131 Padova, Italy}
            
		\vspace{15pt}

			$^{c}${\it   Physics Division, National Technical University of Athens,\\
            15780 Zografou Campus, Athens, Greece }

 		}

		\vspace{40pt}

		{ABSTRACT}
	\end{center}

	\vspace{10pt}

We derive the 3D N=1 superpotential for the closed string sector 
of type IIB supergravity on toroidal O5 orientifolds with co-calibrated G2 structure and RR background flux. 
We find that such compactifications can provide full closed string moduli stabilization on supersymmetric AdS$_3$ vacua, 
and once we include brane-supersymmetry-breaking we also find indication for the existence of classical 3D de Sitter solutions. 
The latter however are rather difficult to reconcile with the ``shape'' moduli stabilization and flux quantization. 
We also discuss the possibility of achieving scale separation in AdS$_3$ and dS$_3$ vacua, 
but such effects seems to be hindered by the geometric flux quantization.

\bigskip

\vspace*{\fill}
\noindent
\rule{5.8cm}{0.4pt}\\
{\rm \footnotesize E-mails: maxim.emelin@mail.mcgill.ca, fotios.farakos@pd.infn.it, georgiostringas@mail.ntua.gr}

\end{titlepage}


\vskip 0.5cm

\vspace{0.5cm}

\def\thefootnote{\arabic{footnote}}
\setcounter{footnote}{0}

\baselineskip 6 mm






\tableofcontents

\section{Introduction}

One of the most drastic twists in the study and interpretation of string flux compactifications (see e.g. \cite{Becker:2007zj}) 
is the notion of the swampland \cite{Vafa:2005ui,Ooguri:2006in,ArkaniHamed:2006dz}. 
The central proposal of the swampland program is that the various common properties of string theory vacua 
can be interpreted as manifestations of the underlying rules that any theory of quantum gravity should adhere to, 
and not special instances of our inability to find solutions with different properties. 
This means that behind generic flux compactifications there exist underlying fundamental quantum gravity rules, 
that govern their properties and so cannot be violated. 
Since these rules are unproven they are proposed as conjectures, 
which are in turn tested on the only available quantum gravity theory we know, 
that is string theory. 
If such conjectures seem to hold in string theory then one should search for the 
underlying reasons for such behavior rooted within quantum gravity. 
This means that the notion of the swampland is not only a tool that allows us to facilitate the study of the 
vast string theory landscape, 
but is essentially deeper than that, 
as it applies to any quantum gravity theory and to any string flux compactification 
- relevant to our universe or not. 
For reviews on the swampland program see for example \cite{Palti:2019pca,Andriot:2020lea,vanBeest:2021lhn}.

The generality of the swampland conjectures means they should also apply to flux compactifications 
with any number of external dimensions, unless of course there are quantum gravity reasons 
to expect a specific dimensional dependence. 
Conversely, 
if the qualitative properties of flux compactifications depend on the external dimensions, 
then from the perspective of the swampland this means that some aspects of quantum gravity are intrinsically different across dimensions. 
Therefore, 
string flux compactifications 
down to dimensions different than four 
are a valuable resource for our understanding of the swampland. 
In particular, three-dimensional compactifications are especially interesting for a number of reasons. 
Firstly they are dual to two-dimensional field theories living on the boundary 
and in the case of supersymmetric AdS vacua, 
this means two-dimensional supersymmetric CFTs (for a sample of recent work see e.g. \cite{Eberhardt:2017uup,Dibitetto:2018ftj,Zacarias:2021pfz,Legramandi:2020txf,Faedo:2020lyw,Passias:2020ubv}). 
As a result the properties of such vacua can be cross-checked with 2D CFT methods. 
Secondly, from a technical point of view, 
the field content of a 3D flux compactification is considerably simpler than the 4D counter-parts which allows 
to perform a more thorough study of such vacua \cite{Farakos:2020phe,Farakos:2020idt}. 
For example, 
the minimal supersymmetric background in 3D allows half of the number of supersymmetries than 
the minimal 4D background. 
Thirdly, 
since gravitation in 3 dimensions is intrinsically different than four dimensions or beyond, 
the study of the 3D swampland offers a unique ground to test the dependence of the 
conjectures on the dimensions of the external space.

For the reasons outlined above, 
in this work we continue the study of Type II string flux compactifications with three-dimensional external space 
and minimal supersymmetry, that was initiated in \cite{Farakos:2020phe,Farakos:2020idt}. 
As shown in those works, Type II on G2 holonomy manifolds 
has offered the possibility to scrutinize the swampland conjectures. 
Here we pursue this direction further by working instead with manifolds with G2-structure. 
There is an extended bibliography on flux compactifications with G2-strucutre, 
for example 4D vacua have been studied in  \cite{Bilal:2001an,Beasley:2002db,DallAgata:2005zlf,Derendinger:2014wwa,Danielsson:2014ria,Andriolo:2018yrz} 
and 3D vacua of Heterotic strings have been studied for example in \cite{delaOssa:2017gjq,delaOssa:2019jsx}. 
The simplest deviation from G2 holonomy is co-calibrated G2-structures, which will be our main focus.
As we will see, 
since we want to reduce the amount of preserved supersymmetry to the minimum, i.e. N=1 in 3D, 
Type IIB offers a preferred framework, compared to IIA, due to the fact that O5/O9 planes are naturally compatible with the co-calibrated G2-structure.

In the rest of this work we first present the background geometry and then perform a direct dimensional 
reduction of IIB supergravity on spaces with co-calibrated G2-structure. 
We then work out the 3D superpotential and verify our findings via an appropriate S-duality. 
As an application, we study moduli stabilization within the 3D EFT framework which yields SUSY-AdS$_3$ vacua. 
We also consider a related setup involving brane-supersymmetry-breaking (BSB) that has been developed 
and studied for example in 
\cite{Antoniadis:1999xk,Angelantonj:1999ms,Dudas:2000nv,Pradisi:2001yv,Mourad:2017rrl,Basile:2018irz,Basile:2020mpt}, 
which allows for non-SUSY Anti-de Sitter as well as de Sitter vacua. 
For the de Sitter vacua we study first only the volume-dilaton sector 
and we see that stable critical points are allowed, 
however when we also switch-on the shape moduli we find that 
they pose a threat to stabilization in de Sitter. 
In addition, 
in all cases we find that scale-separation 
is in tension with certain quantization conditions.

\section{Type IIB on toroidal orbifolds}

\subsection{Integrable G2 structures}

In this subsection we discuss the basic features of the seven dimensional internal space $X$ to be used in our compactifications.
The G2 structure is characterized by the invariant three-form in an oriented seven dimensional manifold $X$
\be
\label{Phi-rad}
\Phi = e^{127}
-e^{347}
-e^{567} 
+e^{136}
-e^{235}
+e^{145}
+e^{246} \,. 
\ee
For the case of a torus $e^m=r^m dy^m$, 
where $r^m$ stand for the radii of the corresponding cycles 
and $dy^m$ are the orthonormal basis of the internal seven dimensional manifold $X$ 
\begin{align}\label{3formbasis}
    \Phi_i = (+dy^{127}, -dy^{347}, -dy^{567}, +dy^{136}, -dy^{235}, +dy^{145}, +dy^{246}) \; , \;\;\;\;\; i=1,...,7.
\end{align}
The corresponding Hodge dual of the fundamental three-form
\begin{align}\label{4formbasis}
\star\Phi = e^{3456}
-e^{1256}
-e^{1234} 
+e^{2457}
-e^{1467}
+e^{2367}
+e^{1357} \, , 
\end{align}
can be expanded in the basis of 4-forms 
\begin{align}
    \Psi_i = (+dy^{3456}, -dy^{1256}, -dy^{1234}, +dy^{2457}, -dy^{1467}, +dy^{2367}, +dy^{1357}) \; , \;\;\;\;\; i=1,...,7.
\end{align} 
The elements of this four-form basis satisfy $\int \Phi_i\wedge \Psi_j = \delta_{ij}$. 
$\Phi$ can be also used to define the volume as follows 
\begin{align}\label{volume}
    \int \Phi\wedge \star\Phi = 7 \, \text{vol}(X) \,, 
\end{align}
where we are assuming $\int_7 dy^{1234567}=1$. 
Once we recast the torus radii in terms of the scalar moduli $s^i$ that describe deformations of the internal G2 space, 
the invariant 3-form reads 
\be
\label{Phi-exp}
\Phi = \sum_i s^i \Phi_i \,. 
\ee
The exact relation of the $s^i$ to the radii can be found by simply comparing \eqref{Phi-exp} 
to \eqref{Phi-rad} taking into account \eqref{3formbasis}.

The fundamental three-form defines a Riemannian metric and thus a Levi-Civita covariant derivative $\nabla \Phi$ associated to the metric. Generally the three-form is not necessarily covariantly constant 
and then, for an integrable G2, a fully antisymmetric torsion tensor exists 
\begin{align}\label{fulltorsion}
    T_3(\Phi) = \frac{1}{6}W_1 \Phi - \frac{1}{3}W_7\lrcorner (\star\Phi) - \star W_{27} \, , 
\end{align}
which is expressed in terms of $\Phi$ 
and the $W_i$ p-differential forms (we follow the notation of \cite{DallAgata:2005zlf}). 
The latter, which are the so-called ``torsion classes'', 
correspond to the irreducible representations $\textbf{1} \oplus \textbf{7} \oplus \textbf{27}$ of the G2 structure. 
The presence of torsion classes is classified by Fernandez and Gray \cite{FernandezGray}. 
Note that for integrable G2  the torsion which belongs to the $\textbf{14}$ representation must vanish, 
that is $W_{14}=0$, and therefore we have ignored it in the above formulas and in the following as well. 
Then the exterior derivatives on the three form and its dual give the structure equations which can be decomposed 
in terms of the torsion classes 
\be
d \Phi =  W_1\star \Phi -\Phi\wedge W_7+ W_{27} \ , \quad  
d \star\Phi =  \frac{4}{3}\star \Phi\wedge W_7 \, , 
\ee
and satisfy $\Phi\wedge W_{27}=0$. It is important to point out that the non-vanishing torsion $\nabla\Phi \neq 0$ 
signifies a deviation from G2 holonomy and implies a non-vanishing Ricci tensor which is a key feature of this work. 
Previous works \cite{Farakos:2020phe,Farakos:2020idt} studied the case where $\nabla\Phi = 0$ and thus all torsions were 
simultaneously set to zero, the internal space was Ricci flat and the G2 structure group 
was equivalent to the G2 holonomy of the manifold. 
Here instead the form of the Ricci scalar is 
\be
R^{(7)}= -4\star d\star W_7 + \frac{21}{8} W_1^2 + \frac{30}{9}| W_7|^2 
-\frac{1}{2}\vert W_{27}\vert^2  \, . 
\ee

We now restrict our attention to the simplest toroidal example 
where $X=T^7/(Z_2 \times Z_2 \times Z_2)$, 
presented in \cite{DallAgata:2005zlf,Joyce}, 
which is a seven torus orbifolded under the action of the following $Z_2$ involutions 
\be
\begin{aligned}\label{Z2s}
\Theta_\alpha : (y^1, \dots, y^7 ) & \to (-y^1, -y^2, -y^3, -y^4, +y^5, +y^6, +y^7) \, , 
\\
\Theta_\beta : (y^1, \dots, y^7 ) & \to (-y^1, -y^2, +y^3, +y^4, -y^5, -y^6, +y^7) \, ,
\\
\Theta_\gamma : (y^1, \dots, y^7 ) & \to (-y^1, +y^2, -y^3, +y^4, -y^5, +y^6, -y^7) \, . 
\end{aligned}
\ee
The orbifold group is $\Gamma=\{\Theta_\alpha,\Theta_\beta,\Theta_\gamma\}$ 
and therefore one automatically has to take into account the combined involutions 
\be
\begin{aligned}
\Theta_{\alpha}\Theta_{\beta} : y^i & \to (+y^1, +y^2, -y^3, -y^4, -y^5, -y^6, +y^7) \, , 
\\
\Theta_{\beta}\Theta_{\gamma} : y^i & \to (-y^1, -y^2, +y^3, +y^4, -y^5, -y^6, +y^7) \, ,
\\
\Theta_{\gamma}\Theta_{\alpha} : y^i & \to (+y^1, -y^2, +y^3, -y^4, -y^5, +y^6, -y^7) \, , 
\\
\Theta_{\alpha}\Theta_{\beta}\Theta_{\gamma} : y^i & \to (-y^1, +y^2, +y^3, -y^4, +y^5, -y^6, -y^7) \, . 
\end{aligned}
\ee
Note that the maps $\Theta$ commute, they square to the identity, 
preserve the both the calibration $\Phi$ and its Hodge dual. 
The above involutions allow us to twist the torus by introducing non-zero metric fluxes $\tau^i_{jk}\neq 0$ 
and thus deviate to a G2-structure manifold. 
We follow the steps of \cite{DallAgata:2005zlf,Derendinger:2014wwa} 
and twist the torus {\it \`a la} Scherk--Schwarz \cite{Scherk:1979zr}. 
For the twisted torus one replaces the straight differential forms $dy^i$ with twisted 1-forms $dy^i \to \eta^i$
which satisfy the Maurer--Cartan equation 
\be
d\eta^i = \frac{1}{2}\tau^{i}_{jk}\eta^j\wedge \eta^k \,. 
\ee
This means we now have twisted vielbeins 
\be
e^i = r^i \eta^i \, ,
\ee 
and in particular in the previous expressions one does the replacements $dy^{ijk} \to \eta^{ijk}$ and $dy^{ijkl} \to \eta^{ijkl}$. 
From the Scherk--Schwarz reduction the geometric flux is constrained by 
\be 
\label{TAU-TAU}
\tau^i_{ji} = 0 \ , \quad \tau^l_{[ij}\tau^{m}_{k]l}=0  \, . 
\ee
These conditions restrict the possible $\tau^l_{ij}$ values. 
In particular the specific orbifold group further projects out the torsion classes $W_7$ 
and therefore the exterior derivatives become 
\be
\begin{aligned}
& d \Phi  =  W_1\star \Phi + W_{27} \, ,  \\
& d \star\Phi  = 0 \, , 
\end{aligned}
\ee
which is the case of \textit{co-calibrated} G2-structures due to the closure of $\star\Phi$. 
This actually happens because the eliminated torsion classes, which were one- and two-forms, 
were not invariant under the orbifold action. 
The Betti numbers, 
which depend on the presence of $W_i$, 
now coincide with those of the G2 holonomy case 
\begin{align}
\label{BETTI}
    b_1(X)=0 \, , \ b_2(X)=0 \,, \ b_3(X)=7. 
\end{align} 
This means that the torsion class $W_{27}$ can be expanded in the fundamental basis $\Psi_i$, 
which will be important for our calculations later.

For later use, 
we would also like to recall the useful Hodge dual expressions for the co-calibrated G2-structure 
\be
\star \Phi = \sum_i \frac{\text{vol}(X)}{s^i} \Psi_i \ , \quad \star \Phi_i = \frac{\text{vol}(X)}{(s^i)^2} \Psi_i \,. 
\ee
In addition, following \cite{DallAgata:2005zlf}, we can also define the geometric flux matrix 
\be
{\cal M}_{ij} = \int_7 \Phi_i \wedge d \Phi_j \, , 
\ee
such that $d \Phi_i = \sum_j {\cal M}_{ij} \Psi_j$. 
The values of ${\cal M}_{ij}$ depend on the coefficients $\tau^i_{jk}$ in the following way 
\be\label{mijmatrix}
{\cal M}_{ij} = \frac12 \begin{pmatrix}
  0 & -\tau^{7}_{5,6}  & -\tau^{7}_{3,4}  & +\tau^{1}_{4,5}  & +\tau^{2}_{4,6} & +\tau^{1}_{3,6}  & -\tau^2_{3,5} \\ 
  -\tau^{7}_{5,6} & 0 & +\tau^7_{1,2} & +\tau^3_{2,5} & -\tau^3_{1,6} & -\tau^4_{2,6} & -\tau^4_{1,5} \\
  -\tau^{7}_{3,4} & +\tau^7_{1,2} & 0 & +\tau^6_{2,4} & +\tau^5_{1,4} & -\tau^5_{2,3} & +\tau^6_{1,3} \\
  +\tau^1_{4,5} & +\tau^3_{2,5} & +\tau^6_{2,4} & 0 & -\tau^3_{4,7} & +\tau^1_{2,7} & +\tau^6_{5,7} \\
  +\tau^2_{4,6} & -\tau^3_{1,6} & +\tau^5_{1,4} & -\tau^3_{4,7} & 0 & -\tau^5_{6,7} & -\tau^2_{1,7} \\
  +\tau^1_{3,6} & -\tau^4_{2,6} & -\tau^5_{2,3} & +\tau^1_{2,7} & -\tau^5_{6,7} & 0 & +\tau^4_{3,7} \\
  -\tau^2_{3,5} & -\tau^4_{1,5} & +\tau^6_{1,3} & +\tau^6_{5,7} & -\tau^2_{1,7} & +\tau^4_{3,7} & 0 
\end{pmatrix} \, . 
\ee
With the use of this matrix one can show that  
\be
W_1 = \frac{1}{7 \text{vol}(X)} \sum_{i,j} s^i  {\cal M}_{ij} s^j 
= \frac{1}{7} \left( \prod_k s^k \right)^{-1/3} \sum_{i,j} s^i  {\cal M}_{ij} s^j \,, 
\ee
which gives an exact expression for the $W_1$ torsion class in terms of the moduli $s^i$ and the geometric fluxes.

\subsection{O5-planes}

Now we would like to turn the discussion to the relation of the orientifolds and the orbifold group $\Gamma$. 
In \cite{Farakos:2020phe} we worked with Type IIA and space-filling O2-planes, 
however now the presence of both torsion and O2-planes is forbidden by the Maurer--Cartan equation, 
which automatically sets the structure constants to zero and brings us back to G2 holonomy. 
Here instead we will focus on Type IIB where we can have O3, O5, O7 and O9-planes. 
Due to the lack of one-cycles and five-cycles in co-calibrated G2 
the O3s and O7s are excluded in our setup. 
Therefore we focus on the O5s and O9s which as we will see fit nicely within the co-calibrated G2 setup. 
For the O5s, 
we choose the source current to be proportional to the associated four-form 
\begin{align}
\label{J4O5}
J_4(O5)\sim\sum_i \Psi_i \, , 
\end{align}
i.e. O5-planes wrap 3-cycles inside the G2 space that need to be calibrated in a supersymmetric manner, 
if one wants to achieve a 3D supergravity. 
In this way the O5 involutions match with the orbifold group $\Gamma$. 
Then the O5s sit at the fixed points of the involutions \eqref{Z2s} and 
their positions are shown in the following diagram 
\begin{align}
\label{TOTALorientifolds}
\begin{pmatrix} 
&{\rm O}5_{\alpha}:\quad & - & - & - & - & \times & \times & \times  \\
&{\rm O}5_{\beta} :\quad & - & - & \times & \times & - & - & \times  \\
&{\rm O}5_{\gamma}:\quad &- & \times & - & \times & - & \times & -  \\
&{\rm O}5_{\alpha\beta}:\quad &\times & \times & - & - & - & - & \times  \\
&{\rm O}5_{\beta\gamma}:\quad & \times & - & - & \times & \times & - & -  \\
&{\rm O}5_{\gamma\alpha}:\quad & \times & - & \times & - & - & \times & -  \\
&{\rm O}5_{\alpha\beta\gamma}:\quad & - & \times & \times & - & \times & - & -  \\
\end{pmatrix} \, . 
\end{align}
The ``$\times$'' symbol denotes the directions on the internal $X_7$  manifold spanned by the O5 worldvolume, 
while the ``$-$'' denotes the ``localized'' (modulo smearing) positions (i.e. $0$ and $1/2$) of the O5-planes, 
related to the wrapped cycles by Hodge duality. 
This gives the following currents 
\be
j_{\alpha} = - e^{1234} \ , \quad  j_{\beta} = - e^{1256} \ , \quad  j_{\gamma} = e^{1357} \,,
\ee 
and also $j_{\alpha\beta}$, etc. 
One can also deduce the smeared contribution of the O5s to the three-dimensional effective action. 
For example, for the $\alpha\beta$ 3-cycle 
\be
\frac{j_{\alpha\beta}}{\text{vol}(\alpha\beta)_4}=\frac{e^{3456}}{r^3r^4r^5r^6}=\Psi_1 \, , 
\ee
we would have 
\be
\label{SO5-example}
S_{O5}\sim\int_{O5_{\alpha\beta}} \sqrt{-g_6} = \int_{3} \sqrt{-g_3}\int_{3-\text{cycle}}\sqrt{g_3} = \int_{3} \sqrt{-g_3}\int_{\Phi^1}\star j_{\alpha\beta} = \int_{3} \sqrt{-g_3}s^1 \,. 
\ee 
We will give the exact and more compact form of \eqref{SO5-example} in the next section.

We see that the O5s are compatible with the G2 involutions. 
However, 
we should also ask that when we combine the O5 involutions $\sigma(O5)$ with the G2 then 
the generated involution is also due to a physical object. 
In other words we always ask the images of O-planes to be O-planes. 
For example if we take 
\be
\sigma(O5_{\alpha}) \Theta_{\beta} : \eta^i \, \equiv \, \sigma(O5_{\alpha\beta}) : \eta^i \, , 
\ee
we verify that the web of O5-planes is generated. 
Now, 
there are six non-trivial combinations that generate all the O5s even if we 
assumed the existence of only one of them, 
but there is also a combination that leads to the identity. 
That is 
\be
\sigma(O5_{\alpha}) \Theta_{\alpha} : \eta^i \to \eta^i \,, 
\ee
which has to be identified as an involution arising from an O-plane. 
Clearly the only candidate is the O9-plane, 
which is also 10D space-filling, and can be mutually supersymmetric with the O5s. 
To this end we have 
\be
\sigma(O5_{\alpha}) \Theta_{\alpha} : \eta^i \, \equiv  \, \sigma(O9) \eta^i \,. 
\ee
This means our setup really resides in type I string theory. 
Naturally, the configuration must also include a suitable number of D9-branes, 
resulting in an open string sector, which we will largely ignore in this work. 
Alternatively, we could consider a similar setup with O5s and O9$^+$ 
resulting in the brane-supersymmetry-breaking scenario \cite{Mourad:2017rrl}. 
We will return to this case in section \ref{BSB}.

\subsection{The scalar potential from 10D}

Since we plan to perform a dimensional reduction of Type IIB on a background that includes 
O-planes we now discuss the possible background fluxes we can introduce and the 
field content of the 3D effective theory. 
We will need to discuss only the bosonic sector as the fermionic sector is fixed by supersymmetry. 
The latter, 
because of the O5-planes on top of the G2, is left with only two independent Killing spinors, 
that is we have 3D N=1 local supersymmetry. 
The gravity sector will essentially include the 3D external metric $g_{\mu \nu}$ and the seven $s^i$ moduli 
that parametrize the twisted torus radii. 
We will further split them into the overall volume modulus $v$ and the unit-volume deformations $\tilde s^i$ 
(which we will often refer to as shape moduli). 
These seven moduli, together with the dilaton $\phi$, 
form the full set of eight real scalar moduli that will enter the 3D theory. 
Indeed, 
other scalar moduli would only arise from the reduction of the RR fields or the NS two-form and 
we will outline now why they are not a part of the 3D effective theory.

We will follow \cite{VanRiet:2011yc} for the rules of the parities of the various fields, 
and we focus explicitly on the parities under the O5s. 
First we note that the $H_3$ has to be odd and so does the $H_7$. 
Since there is no odd 3-form basis to expand $H_3$ on (or a 7-form to expand $H_7$), the $H$ flux has to vanish. 
In addition the co-calibrated toroidal G2 has no one- or two-cycles (the Betti numbers are given by \eqref{BETTI}) 
and so the 3D fluctuations of the $B_2$ NS gauge two-form are truncated. 
Now we turn to the RR sector. 
The $C_0$ RR field, which would be a scalar, 
is odd under the O5-plane and so its 3D fluctuations are truncated. 
The $F_1$ flux cannot be part of the background as there are no one-cycles. 
The $C_2$ RR field is even under the O5, however, 
due to the lack of one- or two-cycles it does not give 
rise to vector or scalar fluctuations in 3D. 
In addition, 
2-forms in 3D are auxiliary fields and so they only contribute via their 3-form background flux. 
Indeed, 
the 3-form RR flux $F_3$ can have non-vanishing values. 
The $F_3$ is even under the O5 parity and therefore can be expanded on the basis of the even forms $\Phi_i$, 
whereas the $F_7$, 
which is also even under O5, will just be proportional to the volume form of the internal space. 
Finally the $C_4$ is odd under O5 parity and since there are no odd 3- or 4-cycles 
it does not give rise to any 3D fluctuations. 
In addition, its $F_5$ flux would need to be expanded in a basis of odd 5-forms which do not exist 
in the co-calibrated G2. 
As a result $F_5$ (and $C_4$) are completely truncated. 
This verifies that the 3D supergravity will only have 
the seven radii of the torus together with the dilaton $\phi$ as scalar moduli.

The (pseudo) action for the Type IIB supergravity in the Einstein frame is given by 
the sum of the NSNS and the RR parts bellow 
\be
\begin{aligned}
&S_{\text{NS}}=\frac{1}{2\kappa^2}\int d^{10}x\sqrt{-g}\Big(R-\frac{1}{2}\partial_{M}\phi\partial^{M}\phi 
- \frac{1}{2}e^{-\phi}\vert H_3 \vert^2 \Big) 
\, ,  \\ 
&S_{\text{RR}}=\frac{1}{2\kappa^2}\int d^{10}x\sqrt{-g}\Big( - \frac{1}{2}\sum_n e^{\frac{5-n}{2}\phi}\vert F_n \vert^2 \Big) \, , 
\end{aligned}
\ee
where n runs over 1, 3 and 5. 
The Born--Infeld part of either of the Dp-brane or Op-plane actions in the Einstein frame is
\begin{align}
S_{\text{loc}}= \mu_p \int e^{\frac{p-3}{4}\phi}\sqrt{-g_{p+1}} \,,  
\end{align}
where $\mu_p>0$ for O-planes and $\mu_p<0$ for D-branes. 
We will give momentarily the details about the Bianchi identities that are related to the couplings of these 
objects to the RR fields.

We can now perform a direct dimensional reduction down to 3D. 
In 10D Einstein frame,  our reduction Ansatz for the metric is 
\be
\label{10DmetricV}
ds^2_{10} =  e^{2\alpha v} ds^2_3 + e^{2\beta v}\widetilde{ds}^2_7\,,
\ee
where $v$ is a 3D scalar that accounts for the compactification volume and hence $\widetilde{ds}^2_7$ is the metric on a unit-volume G2 space. 
The world indices then break into external and internal respectively as $M=(\mu,m)$, 
where $\mu=0,1,2$ and $m=1,\dots,7$. 
The potential energy contributions to the three-dimensional action, 
that arise after the compactification from the ten-dimensional action considering the reduction Ansatz, 
are 
\be
\begin{aligned}
    V_R &= - \tilde R^{(7)} \, e^{-2\beta v} e^{2\alpha v} \, ,  \\
    V_{\text{flux}} &= \frac{1}{2} \vert F_q \vert^2 \, e^{\frac{5-q}{2}\phi} e^{2\beta v(\frac{7}{2}-q)}e^{3\alpha v} \, ,  \\
    V_{\text{D}_{\text{p}}/\text{O}_{\text{p}}} &= -\mu_p  \, e^{\frac{p-3}{4}\phi} \, e^{2\beta v(\frac{2p-11}{4})}e^{\frac{5}{2}\alpha v} \label{DPOP} \, , 
\end{aligned}
\ee
where the Ricci scalar of the co-calibrated G2 internal space is
\be
\tilde R^{(7)}=  \frac{21}{8} \tilde W_1^2 -\frac{1}{2}\vert \tilde W_{27}\vert^2  \, , 
\ee
and
\be
\vert F_q \vert^2 = F_q \wedge \tilde \star F_q = \sqrt{\tilde g} \frac1{q!} F_{m_1 \dots m_q} F^{m_1 \dots m_q} \,. 
\ee
Note however that $\sqrt{\tilde g} = 1$. 
Now with the specific choice of numbers 
\be\label{abrelation}
\alpha^2 = 7/16 \ , \quad \beta = -\frac{1}{4 \sqrt{7}} \ , \quad -7\beta =\alpha \, , 
\ee
we find canonical kinetic terms for the volume-dilaton in three dimensions 
\be
\label{kinetic+V}
e^{-1}\mathcal{L} = R_3 - \tfrac{1}{2}(\partial v)^2 - \tfrac{1}{2}(\partial \phi)^2 - V + \dots 
\ee
Let us now recall that the only background RR fluxes that we can switch on due to the O5 truncation 
(or the parity restrictions) 
are given by 
\be
\label{F7F3-back} 
F_7 
= - {\cal G} \, dy^{1234567}  \ , \quad F_3 = \sum_i f^i \Phi_i \, ,  
\ee
which are consistent with tadpole cancellation, since $H_3=0$ and 
\be
d F_7 = 0 \ , \quad d F_3 \ne 0 \,. 
\ee 
The latter holds due to the co-calibrated G2-structure which gives rise to torsion. 
Because of that, 
the Bianchi identity for the $F_3$ is satisfied as 
\be
\label{WithoutD5s}
d F_3 = - \mu_{O5} J_4 (O5) \ , \quad \text{no D5-branes} \, . 
\ee
As a result such background does not require D5s for the cancellation of the O5 source even though 
the NS H-flux is identically vanishing. 
In the presence of D5s the Bianchi identity (again for $H_3=0$) becomes 
\be
\label{WithD5s}
d F_3 = - \mu_{O5} \, J_4(O5) - \mu_{D5i} \, J_4(D5i) \,, 
\ee 
where $\mu_{O5} > 0$ and $\mu_{D5i} < 0$, 
and we readily identify the O5-plane/D5-brane charges with their tension (up to the dilaton factors) 
because they are supersymmetric BPS objects. 
Here we indicate with $J_4(D5i)$ the source current for the D5s wrapping the $i$-th 4-cycle. 
Even though the tadpole cancellation is seemingly possible without the use of D5s, 
namely as in \eqref{WithoutD5s}, 
as we will see when we turn to explicit examples we will often need to use \eqref{WithD5s}. 
For a review of the IIB ingredients we have used here see e.g. \cite{Blaback:2010sj}.

One may be worried about $dF_3 \ne 0$ because it implies the existence of magnetic sources 
and that the $F_3$ is not closed any more, 
which means that there can be inconsistencies if in our theory a bare $C_2$ RR field also appears. 
However, it is important to appreciate that IIB supergravity does not have an honest Lorentz invariant Lagrangian, 
and as a result the full information of the consistent reduction is captured by the 10D equations of motion 
and the 10D tadpole conditions. 
In these equations the $C_2$ in fact does not appear, 
it is indeed introduced only after one solves the tadpoles with the condition $dF_3 = 0$. 
Then the effective pseudo-action for IIB can be written down which will also include the $C_2$. 
However, a priori one only has a set of 10D equations of motion and Bianchi equations to solve. 
In our approach we first make sure we satisfy these conditions in the internal space 
and then we look directly at the resultant 3D effective theory.

\section{The 3D N=1 superpotential}

\subsection{The scalar potential of 3D N=1 supergravity}

Now we will construct the superpotential for the 3D N=1 supergravity by matching with 
the scalar potential that we derived from dimensional reduction in the previous parts. 
Since we want to have a 3D Einstein frame with the 
conventional $1/2$ factor in front of the Hilbert--Einstein term, we perform a Weyl rescaling of the external 3D space metric of the form 
\be
\label{3D-WEYL}
g_{\mu \nu}^{\text{from dim. reduction}} = \frac{1}{4} \times g_{\mu\nu}^{\text{in 3D N=1 supergravity}}  \,. 
\ee
This brings the kinetic terms for the scalar moduli and the scalar potential 
from the dimensional reduction to the form 
\be
\label{3D-RESCRESC}
e^{-1} {\cal L}_{\text{kin}} = \frac12 R_3 
- \frac14 \partial v \partial v 
- \frac14 \partial \phi \partial \phi 
- \frac14 \text{vol}(\tilde X)^{-1}\int_7 \Phi_i\wedge\tilde\star\Phi_j \partial \tilde{s}^i\partial \tilde{s}^j 
- \frac18 V^{\text{dim.\,red.}}
\,, 
\ee
where we set the 3D Planck scale to unit and $V^{\text{dim.\,red.}}$ is the scalar potential 
from the direct dimensional reduction. 
We use the tilde ``$\sim$'' symbol to denote that the internal metric used is now the unit-volume one, 
and the internal metric shape moduli ($\tilde s^i$) are also the ones corresponding to the unit-volume. 
We will see momentarily exactly how this works.

In general, 
once we are given the kinetic terms of a 2-derivative 3D N=1 supergravity theory, 
the scalar potential is uniquely fixed by the superpotential, 
the latter being a real function of the scalar multiplets. 
In contrast to 4D N=1 here the superfields are real and so the superpotential is also real. 
In addition the scalar manifold is only required to be Riemannian and there is 
no prepotential required to generate it. 
To be precise, 
the scalar sector of 3D N=1 supergravity has the form 
\be
\label{3D-EFT-scalar}
e^{-1} {\cal L}_{\text{scalar}} = \frac12 R_3 - G_{IJ} \partial \varphi^I \partial \varphi^J 
- \left( G^{IJ} P_I P_J - 4 P^2 \right) \, , 
\ee
where $\varphi^I$ are the various real scalar moduli, 
the real function $P(\varphi^I)$ is the superpotential, 
and $P_I = \partial P / \partial \varphi^I$. 
For our setup, the moduli are $\varphi^I = (\tilde s^i, v, \phi)$, 
and therefore the scalar potential has the form 
\be
\label{VfromP}
V = G^{IJ} P_I P_J - 4 P^2 = G^{ij} P_i P_j + 4 P_v^2 + 4 P_{\phi}^2 - 4 P^2 \ , \quad I=i,v,\phi \, ,  
\ee 
where $G^{ij}$ is the inverse of $\frac14 \text{vol}(\tilde X)^{-1}\int_7 \Phi_i\wedge\tilde\star\Phi_j$ 
and 
\be
P_i = \frac{\partial P}{\partial \tilde s^i} 
\ , \quad P_v = \frac{\partial P}{\partial v} 
\ , \quad P_{\phi} = \frac{\partial P}{\partial \phi} \,.  
\ee 
Note that the $\tilde s^i$ satisfy the condition 
\be
\label{UNIT-VOL}
\text{vol}(\tilde X) = 1 = \left( \prod_i \tilde s^i \right)^{1/3} \,, 
\ee
and therefore for our toroidal orbifold we find explicitly 
\be
\label{Gij}
G_{ij} = \frac14 \text{vol}(\tilde X)^{-1}\int_7 \Phi_i\wedge\tilde\star\Phi_j = \frac{\delta_{ij}}{4 (\tilde s^j)^2} \,. 
\ee
We can solve the condition \eqref{UNIT-VOL} by setting 
\be
\label{ts7tsa}
\tilde s^7 = \prod_{a=1}^6 \frac{1}{\tilde s^a} \,,  
\ee
which we will often invoke throughout this work and in the examples later. 
Then \eqref{Gij} should not be used as the true scalar manifold metric for the $\tilde s^a$ scalars. 
Instead we have to take into account that $\partial_\mu \tilde s^7$ also contains 
derivatives with respect to the $\partial_\mu \tilde s^a$. 
Therefore from \eqref{3D-EFT-scalar} once we take into account \eqref{Gij} and \eqref{ts7tsa} we find 
\be
\label{tGab}
\tilde G_{ab} =  \frac{1 + \delta_{ab}}{4 \, \tilde s^a \tilde s^b} \ , \quad a,b = 1,2,3,4,5,6 \, , 
\ee
such that $G_{ij} \partial \tilde s^i \partial \tilde s^j \equiv \tilde G_{ab} \partial \tilde s^a \partial \tilde s^b$. 
This matrix should be used when one wants to canonically normalize the scalars.

Let us now discuss an important technical point about the way that we evaluate the scalar potential from the superpotential. 
We first take the derivatives of the superpotential with respect to the 
\emph{unrestricted} $\tilde s^i$, and then, after all derivatives have been evaluated, 
we impose the condition \eqref{UNIT-VOL}. 
This procedure is completely consistent because of the specific properties of our superpotential, 
otherwise such procedure would not preserve supersymmetry. 
In particular it was proven in \cite{Farakos:2020phe} that a sufficient condition for doing this is 
\be
\label{GPi}
G^{ij} P_i  \int \Phi_j \wedge \tilde \star \tilde \Phi = 0 \,, 
\ee 
which we will see is always satisfied  by our superpotential. 
In particular, 
when \eqref{GPi} holds then the condition \eqref{UNIT-VOL} is fully supersymmetric in the sense that 
once we act on the latter with a supersymmetry transformation it also eliminates the 
fermion superpartner of the extra scalar and also the extra auxiliary field. 
In fact the condition \eqref{GPi} also guarantees that 
\be
\label{P=P}
P (s^i) \equiv P(\tilde s^i) \,, 
\ee 
where the $s^i$ are the original seven moduli of the G2 that describe the internal metric deformations 
\be\label{sirelation}
s^i = \text{vol}(X)^{3/7}  \tilde s^i = e^{3 \beta v} \tilde s^i \,. 
\ee 
As a result, due to \eqref{P=P}, we can present our superpotentials in terms of $s^i$ instead of $\tilde s^i$ to avoid cluttering, 
when possible, 
and without jeopardizing the result. 
However when we act with $\tilde s^i$ derivatives we have to recast them in terms of $\tilde s^i$ first and then take derivatives. 
We also stress once more that because we performed a 3D Weyl rescaling after the dimensional reduction we will have 
\be\label{rescaled}
V^{\eqref{VfromP} \ \text{from 3D N=1 superpotential}} \ = \ \frac18 \times V^{\text{from dim. reduction}} \, . 
\ee
This means we multiply the scalar potential found from the dimensional reduction with $1/8$ 
to match to the scalar potential we get from the superpotential calculation. 
In this way the supersymmetric theory \eqref{3D-EFT-scalar} will agree with \eqref{3D-RESCRESC}. 
More details about the 3D N=1 supergravity can be found in \cite{Farakos:2020phe}, 
and a more detailed account of its properties can be found in \cite{Becker:2003wb,Buchbinder:2017qls}. 
Finally, 
the quadratic gravitino sector has the form 
\be
\label{gravitino}
e^{-1} {\cal L}_{3/2} = - \frac12 \overline \psi_\mu \gamma^{\mu \nu \rho} D_{\nu} \psi_\rho 
- \frac12 P \, \overline \psi_\mu \gamma^{\mu \nu} \psi_\nu  \, , 
\ee
from which we can verify that for SUSY-AdS$_{3}$ we have 
$\langle P_I \rangle = 0$ and $m_{3/2} = P = (2 L_{\rm AdS})^{-1}$ 
as dictated by the universal properties of supergravity \cite{Freedman:2012zz}.

In the rest of this section we will present the total superpotential $P$ in three steps: 
First we will present the superpotential that corresponds to the internal curvature contribution, 
then the one that corresponds to the $F_3$ flux, 
and then the one that corresponds to the $F_7$ flux. 
Since we essentially guess these contributions, 
we only need to check them by matching with the respective terms in the dimensional reduction scalar potential. 
Moreover, 
we will see that these three contributions to the superpotential can be combined without 
generating additional terms in the scalar potential, except one, which reproduces precisely 
the scalar potential term from the calibrated and smeared O5-planes (and possibly D5-branes). 
This cross-term is generated from the mixing of the internal curvature superpotential with the 
superpotential for $F_3$. 
Crucially it is the $F_3$ that is used in the tadpole cancellation conditions in the 10D supergravity 
and relates directly to the consistent incorporation of the O5-planes. 
This means that 3D N=1 supergravity is somehow aware of the 10D tadpole cancellation conditions 
and automatically takes them into account.

\subsection{Superpotential from geometric flux}

The superpotential for the internal curvature, 
i.e. the geometric flux, is 
\begin{align}
\label{SuperR}
    P^R=\frac{1}{16}e^{-8\beta v}\int \Phi\wedge d\Phi \, \text{vol}(X)^{-\frac{6}{7}} \,. 
\end{align} 
From \eqref{SuperR} we directly see that 
\be
\label{PRRR}
P^R_v = -\frac{\beta}{2} P^R \ , \quad P^R_{\phi} = 0 \,. 
\ee
For the derivatives with respect to $\tilde s^i$ we have 
\be
\begin{aligned}
P^R_i = &\ \frac{e^{-8\beta v}}{16} \Big( \int \Phi_i \wedge d \tilde \Phi \, \text{vol}(\tilde X)^{-\frac{6}{7}} 
+ \int \tilde \Phi \wedge d\Phi_i \, \text{vol}(\tilde X)^{-\frac{6}{7}} \Big) 
\\ 
& - \frac{e^{-8\beta v}}{16} \Big( \frac{6}{7}\int \tilde \Phi \wedge d\tilde \Phi 
\, \text{vol}(\tilde X)^{-\frac{13}{7}} (\text{vol}(\tilde X))_i \Big) \,, 
\end{aligned}
\ee
where $\tilde \Phi = \tilde s^i \Phi_i$ and we have $d(A_3 \wedge B_3) = dA_3 \wedge B_3 - A_3 \wedge dB_3$. 
Then we use the following identities 
\be
\label{some-ident}
(\text{vol}(\tilde X))_i = \left( \frac17 \int \tilde \Phi \wedge \tilde \star \tilde \Phi \right)_i 
= \frac13 \int \tilde \Phi \wedge \tilde \star \Phi_i 
\ , \quad 
\Phi_i \wedge d\tilde \Phi = \tilde \Phi \wedge d\Phi_i \, , 
\ee
which bring the derivative of the superpotential with respect to $\tilde s^i$ to the form 
\be
\label{PRi}
P^R_i = \frac{1}{8}e^{-8\beta v} \, 
\Big(\int \Phi_i \wedge d\tilde \Phi 
- \frac{\int \tilde \star \Phi_i\wedge \tilde \Phi}{\int \tilde \star \tilde \Phi \wedge \tilde \Phi} 
\int \tilde \Phi \wedge d\tilde \Phi \Big) \, 
\text{vol}(\tilde X)^{-\frac{6}{7}} \,. 
\ee 
We stress that the second formula in \eqref{some-ident} is not an integration by parts, 
rather it is an exact algebraic identity. 
From this we can also deduce 
\be
P^R_i = \frac{1}{8}e^{-8\beta v} \, \int_7 \Phi_i \wedge \tilde W_{27} 
\, \text{vol}(\tilde X)^{-\frac{6}{7}} \,, 
\ee
where $\tilde W_{27} = W_{27} (\tilde s^i)$ (and we will similarly use $\tilde W_{1}$ shortly). 
This equation means that $W_{27}$ sources the supersymmetry breaking due to torsion 
and that if it vanishes then the $P^R_i$ vanish identically.  
A more extensive account of the properties we used here can be found in \cite{Farakos:2020phe}, 
from which one can also prove that \eqref{PRi} satisfies \eqref{GPi}.  
Now we insert the three pieces $P^R_{i,v, \phi}$ into the formula \eqref{VfromP} and obtain
\begin{align}
\label{VRsusy}
    V^R \Big{|}_{\text{vol}(\tilde X)=1} 
    = \frac{1}{8}e^{-16\beta v} 
    \Big( -\frac{21}{8} \tilde W^2_1+\frac{1}{2}\vert \tilde W_{27}\vert^2 \Big) 
    = - \frac{\tilde R_7}{8} e^{-16\beta v}  \,, 
\end{align}
which is exactly the desired result. 
Note that this corresponds to the Ricci scalar found in \cite{Bryant:2005mz}, 
but here we write it in the notation of \cite{DallAgata:2005zlf}, 
and also it is automatically multiplied by the correct volume prefactor that appears from the dimensional reduction. 
As a technical remark, in deriving \eqref{VRsusy} we needed to contract \eqref{PRi} with $G^{ij}$, 
and to do this we have used in various instances the identity 
\be
\label{GijAB}
G^{ij}  \int \Phi_i \wedge A  \int \Phi_j \wedge B 
= \frac{4}{7} \int \tilde \Phi \wedge \tilde \star \tilde \Phi \int \tilde \star A \wedge B \, , 
\ee
which can be checked by expanding $A = \tilde \star \Phi_i A^i$ and $B = \tilde \star \Phi_i B^i$ 
(see e.g. \cite{Beasley:2002db}). 
For example, this identity was used to derive 
\begin{align}
\nonumber
G^{ij}\int \Phi_i\wedge d\tilde{\Phi} \int \Phi_j \wedge d\tilde{\Phi}
&=\frac{4}{7} \int \tilde{\Phi}\wedge \tilde{\star} \tilde{\Phi}\int \tilde{\star} d\tilde{\Phi} \wedge d\tilde{\Phi} \\
& = 4\Big(\tilde W_1^2\int \tilde{\Phi}\wedge \tilde{\star} \tilde{\Phi} 
+ \int \tilde W_{27}\wedge\tilde{\star} \tilde W_{27} \Big) \nonumber \\
& = 28 \tilde W_1^2 + 4|\tilde W_{27}|^2 \,. 
\end{align}
Here of course we have considered that $d \tilde \Phi$ is expanded in the basis $\Psi_i$, 
otherwise \eqref{GijAB} cannot be used.

\subsection{Superpotential from RR flux}

The superpotential for $F_3$ takes the form\footnote{From the Type I 
perspective one could say that $F_3$ here is in fact $\tilde F_3 = F_3 - \frac14 (\omega_{YM} - \omega_{L})$, 
but we largely ignore here the open string sector in any case.} 
\begin{align}
\label{PF3}
    P^{F3}= - \frac{q}{8}e^{-10\beta v+\frac{\phi}{2}}\int \star\Phi\wedge F_3 \, \text{vol}(X)^{-\frac{4}{7}} \,,  \;\;\;\;\;\; q = \pm 1 \,. 
\end{align}
The role of $q$ and the ambiguity in choosing it is physical and reflects the ambiguity, 
from the 3D supergravity point of view, of introducing O5- or anti-O5-planes. 
In this section we will be working with O5s and we will see shortly how the sign of $q$ can be fixed by matching with the potential from dimensional reduction.
We can again directly evaluate 
\be
\label{PFFF}
P^{F3}_v = -10\beta P^{F3} \ , \quad P^{F3}_\phi = \frac{1}{2} P^{F3} \,, 
\ee
and 
\be
\label{PF3i}
P^{F3}_i= \frac{q}{8}e^{-10\beta v+\frac{\phi}{2}}
\Big(\int \Phi_i \wedge \tilde \star F_3 
-\frac{\int \tilde \star \Phi_i\wedge \tilde \Phi}{\int \tilde \star \tilde \Phi\wedge \tilde \Phi} 
\int \tilde \Phi \wedge \tilde \star F_3 \Big) \, \text{vol}(\tilde X)^{-\frac{4}{7}} \,. 
\ee
Note that \eqref{PF3i} satisfies \eqref{GPi} as anticipated. 
We can also provide an alternative expression that has the form 
\be
P^{F3}_i= \frac{q}{8}e^{-10\beta v+\frac{\phi}{2}} \, 
\int_7 \Phi_i \wedge \tilde \star \pi^{27}(F_3) \, 
\text{vol}(\tilde X)^{-\frac{4}{7}} \,, 
\ee 
where $\pi^{27}(F_3)$ denotes the projection of $F_3$ to the {\bf 27} representation of G2.
Then we insert all these pieces into \eqref{VfromP} 
and through a similar calculation as the one of the previous subsection we find  
\be\label{potentialF3}
V^{F3} \Big{|}_{\text{vol}(\tilde X)=1} 
= \frac{1}{16}e^{-20\beta v + \phi}\int_7 \tilde \star F_3\wedge F_3 \,,  
\ee
which is exactly the contribution to the scalar potential from the RR flux $F_3$. 
Note that we took into account that $q^2=1$ to get to this form. 
In addition, 
we are implicitly assuming $F_3 = f^i \Phi_i$ which means 
\be
d F_3 = f^i d \Phi_i \ \ \ \text{with} \ \ \  d \Phi_i \ne 0 \,, 
\ee
due to torsion. 
However, $d ( \tilde \star F_3) = 0$ because our G2 is co-calibrated, i.e. $d (\tilde \star \Phi_i )= 0$.

Since we have introduced and verified both $P^{R}$ and $P^{F3}$, 
it is now a good time to combine them and uncover the O5-plane/D5-brane contribution to the scalar potential. 
To this end let us take 
\be
\label{R+F3}
P^{R+F3} = P^R + P^{F3} \,. 
\ee
Once we insert \eqref{R+F3} into \eqref{VfromP} we have 
\be
\label{VR+F3}
V^{R+F3} = V^{F3} + V^{R} + 2 G^{IJ} P^R_I P^{F3}_J - 8 P^R P^{F3} \,, 
\ee
where the form of the cross-term is 
\be
\label{SUSY-cross}
\left( 2 G^{IJ} P^R_I P^{F3}_J - 8 P^R P^{F3} \right) \Big{|}_{\text{vol}(\tilde X)=1} 
= \frac{q}8  e^{-18\beta v+ \frac{\phi}{2}} 
\int_{7} \left( \tilde W_1 \, \tilde\star \tilde\Phi \wedge F_3 
+ \tilde W_{27} \wedge F_3 \right) \,. 
\ee
This contribution has to be matched with 
the O5-plane/D5-brane contribution in the effective potential \eqref{DPOP}. 
We will now see how exactly this happens. 
First recall that each of the O5-planes wrap one internal 3-cycle and therefore their currents wrap the dual 4-cycles. 
Therefore for the total O5/D5 contribution we have 
\be
\label{SO5D5}
\begin{aligned}
 S_{O5} + S_{D5s} = & \, \frac18 e^{\phi / 2} \sum_{3-\text{cycles}} ( \mu_{O5} + \mu_{D5i} ) \int_{3D \, \times \, 3-\text{cycle}} \sqrt{-g_6} 
\\
= & \, \frac18  e^{\phi / 2} \sum_{3-\text{cycles}} \left[ \int_{3D \, \times \, 3-\text{cycle}} \sqrt{-g_6} 
\int_{4-\text{cycle}} (\mu_{O5} + \mu_{D5i}) J_4(O5) \right]  \, , 
\end{aligned}
\ee
where the $1/8$ factor comes from the 3D Weyl rescaling \eqref{3D-WEYL}, 
and in going to the second line we have assumed a normalized integration over the 4-cycles 
in the covering space such that 
\be
\int_{ith \, 4-\text{cycle}} J_4(O5) = 1 = \int_{ith \, 4-\text{cycle}} J_4(D5i)  \,. 
\ee 
Let us stress that we are here explicitly ignoring open string moduli related to the D5s, which we assume to be fixed 
on their supersymmetric positions,
otherwise we would have to include them in \eqref{SO5D5} - we leave this interesting development for a future work. 
Now we take into account that for each 6D integral that 
covers the 3D external space and one internal 3-cycle we have 
\be 
\int \sqrt{-g_6}  
= \int_{\text{3D}} \sqrt{-g_3}  \int_{\text{3-cyc.}}  \sqrt{g_3} 
= \int_{\text{3D}} \sqrt{-g_3}  \int_{\text{3-cyc.}}  \Phi 
=e^{3 \beta v + 3 \alpha v}  \int_{\text{3D}} \sqrt{-\tilde g_3}  \int_{\text{3-cyc.}} \tilde \Phi \, , 
\ee
which gives 
\be
S_{O5/D5} =  \frac18 e^{\frac{\phi}{2} + 3 \beta v + 3 \alpha v} \int_{\text{3D}} \sqrt{-\tilde g_3} 
\sum_{\text{3-cycles}} 
\left[
\int_{\text{3-cycle}}  \tilde \Phi  
\int_{\text{4-cycle}} ( \mu_{O5} + \mu_{D5i} )  J_4(O5) \right] \,. 
\ee
To proceed it is instructive to work out the contribution for 
a specific 3-cycle, 
and then recombine all the contributions including the other cycles. 
For example, for $i = 1$, we have
\be
\begin{aligned}
S_{O5/D5(i=1)} =& \, \frac18 e^{-18\beta v + \frac{\phi}{2}} \int_{\text{3D}} \sqrt{-\tilde g_3} 
\, \tilde s^1 \, 
\int_{\text{3-cyc.}} \phi_1   
\int_{\text{4-cyc.}} ( \mu_{O5} + \mu_{D5(i=1)} )  J_4(O5)  
\\
= & \, \frac18  e^{-18\beta v + \frac{\phi}{2}} 
\int_{\text{3D}} \sqrt{-\tilde g_3} 
\int_{7} (\tilde s^1 \, \phi_1) \wedge J_4(O5) (\mu_{O5}+\mu_{D5(i=1)}) 
\\ 
= & - \frac18  e^{-18\beta v + \frac{\phi}{2}} 
\int_{\text{3D}} \sqrt{-\tilde g_3} \, 
\int_{7} (\tilde s^1 \, \phi_1) \wedge dF_3 
\,. 
\end{aligned}
\ee
The last step can be checked by acting with ``$\tilde s^1 \phi_1 \wedge$'' on \eqref{WithD5s}. 
We then perform this procedure for the other six 3-cycles and sum over the results to get the total contribution. 
Taking into account that $\sum_i \tilde s^i \Phi_i = \tilde \Phi$, 
we conclude that 
\be
S_{O5/D5} = \sum_i S_{O5/D5i} =  - \frac18  e^{-18\beta v + \frac{\phi}{2}} 
\int_{\text{3D}} \sqrt{-\tilde g_3} \, 
\int_{7} \tilde \Phi \wedge dF_3 \,.  
\ee
In addition we have that 
\be
\tilde \Phi \wedge dF_3  = d \tilde \Phi \wedge F_3 
= \tilde W_1 \, \tilde\star \tilde\Phi \wedge F_3 
+ \tilde W_{27} \wedge F_3 \, , 
\ee
where the first equality follows from $\tilde \Phi \wedge F_3  \equiv 0$. 
Then we conclude that the total contribution of the smeared O5-planes/D5-branes to the effective 3D potential is 
\be\label{potentialO5}
V^{O5/D5} = \frac18  e^{-18\beta v + \frac{\phi}{2}} 
\int_{7} \left( \tilde W_1 \, \tilde\star \tilde\Phi \wedge F_3 
+ \tilde W_{27} \wedge F_3 \right) \,, 
\ee
which matches exactly with the extra term in \eqref{VR+F3} for 
\be 
q=1 \,. 
\ee
Note that we could in principle split $F_3$ as $F_3=F_{3A}+F_{3B}$ with $F_{3A}\ne0$ 
such that $dF_{3A}=0$ but instead $dF_{3B}=-\mu_{O5}J_4(O5)-\mu_{D5i}J_4(D5i)$, 
which would ``free'' one part of the $F_3$ flux from the tadpole condition.

Finally, for the $F_7$ flux (which is of Freund--Rubin type) the superpotential contribution is 
\be
\label{PF7}
P^{F7} = \frac{1}{8} \, {\cal G} \, e^{-14\beta v - \frac{\phi}{2}} \,, 
\ee
where ${\cal G}$ is a real constant related to the $F_7$ flux \eqref{F7F3-back}. 
Then we evaluate the contribution to the scalar potential which gives 
\be\label{potentialF7}
V^{F7} = \frac{1}{16}  {\cal G}^2 e^{-28\beta v - \phi} \,.
\ee
The superpotential exponential $-14\beta v - \phi/2$ is compatible with the other 
exponentials and does \emph{not} produce any new cross-terms. 
Note that we could have ``$\pm {\cal G}$'' in \eqref{PF7}, 
but only one of the two would correspond to the 10D reduction with $F_7 = -{\cal G} \, dy^{1234567}$, 
the other one would correspond to $F_7 = + {\cal G} \, dy^{1234567}$. 
This ambiguity is fixed by S-duality which chooses the ``$+$'' sign as we will see momentarily. 
We conclude that the full superpotential that describes the dimensional reduction is given by $P^R + P^{F3} + P^{F7}$, 
and reproduces the 3D effective scalar potential (without the brane-supersymmetry-breaking term) 
which one can find by adding the contributions 
(\ref{VRsusy}), (\ref{potentialF3}), (\ref{potentialO5}) and (\ref{potentialF7}), 
and reads 
\be
\begin{aligned}
\label{TOT-tot}
V & = V^R + V^{F3} + V^{O5/D5} + V^{F7} \\
& = - {\rm R}_0(\tilde{s}^i) e^{-16\beta v} +  {\rm F}_0(\tilde{s}^i)e^{-20\beta v + \phi} + {\rm T}_0(\tilde{s}^i)e^{-18\beta v + \frac{\phi}{2}} + {\rm G}_0 e^{-28\beta v - \phi} \,, 
\end{aligned}
\ee
with the coefficients given by 
\be
{\rm R}_0 = \frac{\tilde{R}_7}{8} 
= \frac{1}{64}\left(\sum_{i,j} \tilde{s}^i  {\cal M}_{ij} \tilde{s}^j \right)^2 
- \frac{1}{16}\sum_{i,j}\tilde{s}^i{\cal M}_{ij}\tilde{s}^j\sum_{m}\tilde{s}^m{\cal M}_{mj}\tilde{s}^j  \, , 
\ee
and 
\be
{\rm F}_0 = \frac{1}{16}\sum_i \Big(\frac{f^i}{\tilde{s}^i} \Big)^2 
\ , \quad 
{\rm T}_0 = \frac18  \sum_{k,l} f^l {\cal M}_{lk} \tilde{s}^k 
\ , \quad 
{\rm G}_0 = \frac{{\cal G}^2}{16} \,. 
\ee

\section{Supersymmetric vacua}

\subsection{Supersymmetry cross-check}

As a cross-check of the superpotential of the 3D theory, as well as the overall approach, 
we would like to verify that the 3D vacua that we will find truly describe supersymmetric configurations of the 10D theory. 
Because of the O-plane truncations, the preservation of supersymmetry on our background boils down to 
the supersymmetry Killing equations that arise from Type I string theory with $F_{YM}=0$, 
which can in turn be related to Heterotic string theory via S-duality. 
Earlier work on Heterotic string flux compactifications \cite{delaOssa:2019jsx}, 
has shown that backgrounds with 
\be
H_3^{(\text{HET})} = \frac16 W_1 \Phi - \star W_{27} \ , \quad d \, \phi^{(\text{HET})} = 0 \, , 
\ee
are supersymmetric. 
In other words $H_3^{(\text{HET})}\equiv T(\Phi)$ is identified with the full antisymmetric G2 torsion (\ref{fulltorsion}). 
In \cite{delaOssa:2019jsx} the $W_7$ is also present and relates to the dilaton via 
$d \phi^{(\text{HET})} = 2 W_7$, 
however, the specific orbifolding we use for our twisted torus projects it out, such that we are strictly working with a co-calibrated G2. 
In addition an external component of the $H_3^{(\text{HET})}$ flux is also allowed to be switched on, 
and is also related to the G2 torsion. 
The vacuum condition for the external $H_3^{(\text{HET})}$ flux is 
\be
\label{H-EXT}
H^{(\text{HET - ext})}_{\sigma \lambda \kappa} = - \frac{7 W_1}{6}  e^a_\sigma e^b_\lambda e^c_\kappa \epsilon_{abc} \,, 
\ee 
where $e^a_\sigma$ are the external drei-beins and $\epsilon_{abc}$ is the tangent space full antisymmetric symbol. 
This means $e^a_\sigma e^b_\lambda e^c_\kappa \epsilon_{abc}$ is indeed a tensor. 
Two comments are in order here. 
First, note that in \cite{delaOssa:2019jsx} there is an overall factor $e^n$ that relates the gravitino mass 
to the superpotential via $m_{3/2} = e^n P$. 
Here we have implicitly set it to unit, 
that is we have $n=0$, 
because in our case the gravitino mass is given directly by $m_{3/2} = P$ on supersymmetric AdS 
as seen from \eqref{gravitino}. 
Secondly, in \eqref{H-EXT} we have not performed any additional Weyl rescalings, 
therefore it is still written in the original Heterotic string frame.

To match with Type I string backgrounds we perform an S-duality, 
which for the string frame fields is (see e.g. \cite{Becker:2007zj}) 
\be
H_3^{(\text{HET})} \to F_3 \, , \;\;\;  \phi^{(\text{HET})} \to -\phi 
\, , \;\;\;  g_{MN}^{(\text{HET})} \to e^{-\phi} g_{MN} 
\, , \;\;\;  \Phi \to e^{-3 \phi/2} \Phi 
\, , \;\;\;  \star \Phi \to e^{-2 \phi} \star \Phi 
\,, 
\ee
and also affects the G2 torsion classes as 
\be
W_1 \to e^{\phi / 2} W_1 \ , \quad W_{27} \to e^{-3 \phi/2} W_{27} \,. 
\ee
Therefore on a Type I supersymmetric background one would have the condition 
$e^\phi F_3 = \frac16 W_1 \Phi - \star W_{27}$ for the internal background flux and
the condition $e^\phi F^{(\text{ext})}_3 = - \frac{7 W_1}{6} \sqrt{g_3} \ dt \wedge dx \wedge dz$ for the external background flux, with $t$, $x$ and $z$ being the external space coordinates.
The background dilaton value would still satisfy $d \phi = 0$. 
Then, going to Einstein frame $g_{MN} = e^{\phi/2} g_{MN}^{(\text{E})}$, 
we find 
\be
e^{\frac{\phi}{2}} F_3 = \frac16 W_1 \Phi - \star W_{27} 
\ , \;\quad e^{\frac{\phi}{2}} F^{(\text{ext})}_3 = - \frac{7 W_1}{6} \sqrt{g_3} \ dt \wedge dx \wedge dz 
\ , \;\quad d \, \phi = 0 \, . 
\ee 
From the condition on the external $F_3^{(\text{ext})}$ we find the required vacuum condition on $F_7$ to be 
\be
F_7 = - e^\phi \star_{10} F^{(\text{ext})}_3 =  \frac{7 W_1}{6}\, e^{\frac{\phi}{2}}  \, \sqrt{g_7}  \, dy^{1234567} \,, 
\ee
taking into account that $e^{(5-n)\phi /2}F_n = (-1)^{(n-1)(n-2)/2}\star F_{10-n}$ for $n>5$.
Finally these conditions become 
\be
\label{S-dual-susy} 
e^{\frac{\phi}{2}} F_3 = \frac16 W_1 \Phi - \star W_{27} 
\ , \quad 
e^{-\frac{\phi}{2}} \, {\cal G} = - \frac{7 W_1}{6} \,  \text{vol}(X) \,. 
\ee
We conclude that the conditions \eqref{S-dual-susy} should hold for a supersymmetric vacuum 
of the (Einstein frame) Type I theory and we will re-derive them from our superpotential by requiring $P_I=0$. 
This is a non-trivial cross-check.

We recall that the total superpotential in our setup (we choose $q=1$) reads 
\be 
\begin{aligned}
P = \ & \frac{{\cal G}}{8} e^{-14\beta v -\frac{\phi}{2}}
- \frac{1}{8}e^{-10\beta v + \frac{\phi}{2}}\int \star\Phi\wedge F_3 \, \text{vol}(X)^{-\frac{4}{7}} 
+ \frac{1}{16}e^{-8\beta v}\int \Phi\wedge d\Phi \, \text{vol}(X)^{-\frac{6}{7}} \,. 
\end{aligned}
\ee
We vary $P$ with respect to $v$ and $\phi$ and we require $P_v = 0 = P_\phi$, 
which, after some manipulation, give
\be 
\label{phi-v-SUSY}
{\cal G} \, e^{- \phi / 2} + \frac{13}{7} e^{\phi / 2} \int \star \Phi \wedge F_3 = W_1 \text{vol}(X) 
\, , 
\;\;\;\;\; e^{\phi / 2} \int \star \Phi \wedge F_3  = \frac{11}{7} {\cal G} \, e^{- \phi / 2} + 3 \, W_1 \text{vol}(X) \,. 
\ee
Combining these two equations yields two conditions. 
First we find 
\be
\label{dilaton-stabilization}
\frac67 e^{- \phi / 2} \, {\cal G} = - W_1 \text{vol}(X) \,, 
\ee
which matches exactly with the second condition in \eqref{S-dual-susy}, 
and we also find 
\be
\label{volume-stabilization}
\frac76 \, W_1 \text{vol}(X) =  e^{\phi / 2} \int \star \Phi \wedge F_3 \,, 
\ee
which matches exactly with the first condition in \eqref{S-dual-susy} once we act 
on it with $\int \star \Phi \wedge (\cdot)$, 
taking into account that $\int \star \Phi \wedge \Phi = 7 \text{vol}(X)$. 
Now we take the condition 
\be
\label{tsi-susy}
\frac{\partial P}{\partial \tilde s^i} = 0 \, , 
\ee
where $i=1, \dots, 7$. 
This condition is sufficient to guarantee that $\partial P / \partial \tilde s^a = 0$, 
where $a= 1, \dots 6$ are the true independent $\tilde s^a$ moduli. 
This happens because the unit-volume restriction can be solved as $\tilde s^7 = \prod_{a=1}^6 (\tilde s^a)^{-1}$ 
as we discussed earlier. 
The supersymmetry condition \eqref{tsi-susy} gives 
\be
\label{expanded-pi27-cond}
\int \Phi_i \wedge W_{27} = - e^{\phi / 2} \int \star \Phi_i \wedge \pi^{27}(F_3) \,. 
\ee
Note that $\partial P / \partial \tilde s^a = 0$ would at first sight correspond to only six equations, 
so one can wonder why in \eqref{expanded-pi27-cond} we have seven equations. 
In fact the six $\partial P / \partial \tilde s^a = 0$ equations 
can only be solved once they are expanded in the $\Phi_i$ basis. 
The latter contains seven linearly independent elements and as a result the six $\partial P / \partial \tilde s^a = 0$ equations 
will eventually yield  7 equations (which are exactly \eqref{expanded-pi27-cond}). 
Therefore, since the $\Phi_i$ are a complete basis, we can deduce 
\be
\label{pi27-cond}
\star W_{27} = - e^{\phi / 2} \pi^{27}(F_3) \,. 
\ee
This equation matches exactly with the $\pi^{27}$ part of the first equation in \eqref{S-dual-susy}, 
taking into account that $\pi^{27}(\star W_{27}) \equiv \star W_{27} $. 
Interestingly we see that $W_{27}$ can exist on a supersymmetric vacuum as long as it is cancelled by $\pi^{27}(F_3)$. 
This is in contrast to reductions of M-theory on co-calibrated G2 structures, where supersymmetry requires weak G2 holonomy, 
i.e. $W_{27}=0$ \cite{DallAgata:2005zlf}. 

Let us note at this point that the IIB background we have been considering contains smeared O5-planes, but interestingly, we see an exact match with the Heterotic supersymmetric background. 
This is a non-trivial check for the validity of the effective theory derived from the smeared solution and implies that there should be an underlying full solution 
in IIB where the orientifold sources are localized. 
Indeed, there are instances where the smearing can be ``OK'' 
\cite{Blaback:2010sj,Junghans:2020acz,Baines:2020dmu}.  
We leave the interesting exercise of finding the underlying un-smeared solutions for the future.

\subsection{Conditions for Minkowski and AdS}

We can now examine the possibility of achieving full moduli stabilization 
and determine the required conditions thereof. 
From the conditions on the vacuum, 
that is equations \eqref{S-dual-susy}, 
we find that the vacuum energy of a supersymmetric background is given by 
\be
\label{Vcc}
V\Big{|}_{\text{SUSY}} = - \frac{{\cal G}^2}{16} e^{-\phi} \text{vol}(X)^{-4} \,. 
\ee
From \eqref{Vcc} 
we see that a Minkowski background would require ${\cal G}=0$, 
which from \eqref{S-dual-susy} implies also that $\langle W_1 \rangle = \pi^{1}(F_3) = 0$. 
In other words, 
for a Minkowski vacuum we find the conditions  
\be
\label{pi1-Mink}
\text{SUSY Minkowski}\, : \quad {\cal G} \equiv 0 \ , \quad \langle W_1 \rangle = 0 
= \Big{\langle} \int \Phi \wedge \star F_3 \Big{\rangle} \,. 
\ee
However we can still have non-trivial background flux and $W_{27}$ torsion, 
as long as \eqref{pi27-cond} is satisfied, 
which in fact also tells us that unless $W_{27} \ne 0$ the dilaton is not stabilized. 
Let us now go through the moduli stabilization on Minkowski in more detail. 
Because of the properties of the vacuum conditions it is more 
convenient to work directly with the $s^i$ (and the dilaton of course), 
instead of treating the $\tilde s^a$ and the volume independently. 
Taking into account that $F_3 = f^i \Phi_i$, 
the last equation in \eqref{pi1-Mink} gives 
\be
\label{SFI1}
\sum_i \frac{f^i}{s^i} = 0 \,,  
\ee
which for the moment fixes one of the 8 moduli ($\phi$ and $s^i$). 
Additionally, the condition $W_1 = 0$ gives an equation of the form 
\be
\label{SSM=0} 
W_1 = 0 \quad  \to \quad  \sum_{i,j} s^i {\cal M}_{ij} s^j = 0 \, , 
\ee
which fixes one more of the seven $s^i$ moduli. 
Equation \eqref{pi27-cond} now, 
due to \eqref{pi1-Mink}, 
reduces to 
\be
\label{F3dPhiVAC}
d \Phi = - e^{\phi / 2} \star F_3 \,, 
\ee
where we omit the VEV symbols, since they are implied. 
Then \eqref{F3dPhiVAC} gives 
\be
\label{SFI2}
\sum_i s^i {\cal M}_{ij} = - e^{\phi / 2} \left(\prod_k s^k\right)^{1/3} \frac{f^j}{(s^j)^2} \,, 
\ee
which seemingly amounts to 7 vacuum conditions. 
Note however that \eqref{SFI2} combined with \eqref{SFI1} gives \eqref{SSM=0}, which means 
one of the seven equations of \eqref{SFI2} is already trivially satisfied. 
We therefore conclude that \eqref{SFI2} provides only six additional equations, 
which are however enough to fix the positions of the dilaton and the five remaining $s^i$ moduli. 
Clearly since this is a supersymmetric Minkowski vacuum, 
the absence of tachyonic instabilities is granted from supersymmetry.

Simple Minkowski vacua can be provided by 
the 2-step nilpotent examples of \cite{DallAgata:2005zlf} which in our case read 
\begin{equation}
\label{Mink-Example}
{\small
  {\cal M} = \begin{pmatrix}
    0 & \sigma & \sigma & \sigma & -\sigma & -\sigma & -\sigma  \\
    \sigma & 0 & 0 & 0 & 0 & 0 & 0 \\
     \sigma & 0 & 0 & 0 & 0 & 0 & 0 \\
    \sigma & 0 & 0 & 0 & 0 & 0 & 0  \\
     -\sigma & 0 & 0 & 0 & 0 & 0 & 0 \\
     -\sigma & 0 & 0 & 0 & 0 & 0 & 0 \\
   -\sigma & 0 & 0 & 0 & 0 & 0 & 0   \\
  \end{pmatrix} , 
  }
\end{equation} 
and we choose the values of $F_3$ and ${\cal G}$ to be
\be
f^i = (0,f,f,f,-f,-f,-f) \ , \quad {\cal G} \equiv 0  \,. 
\ee 
The Minkowski vacuum can be found for the values 
\be
\label{VEVS-Mink}
\tilde s^i=1 \ , \quad f = - e^{- \frac{1}{14} (\sqrt{7} v_0 + 7 \phi_0 ) } \sigma \,. 
\ee
We also notice that when the $\tilde s^i$ are fixed on their vacuum values \eqref{VEVS-Mink} 
but $\phi$ and $v$ are left free the scalar potential (\ref{TOT-tot}) takes the form 
\be 
V|_{\tilde s^i =1} = 
\frac38 e^{\frac{4 v}{\sqrt{7}}} \left(e^{\frac{1}{14} (\sqrt{7} v + 7 \phi ) } f + \sigma \right)^2 \,, 
\ee
which is consistent with \eqref{VEVS-Mink} and also indicates the existence of at least one flat direction. 
This should not be confused with no-scale vacua because here $\langle P \rangle =0$. 
Naturally, evaluating the determinant of the mass matrix, we find it to be vanishing.

Let us now turn to AdS supersymmetric vacua. 
Here we allow $P \ne 0$ and therefore we do not have to set ${\cal G}$ to vanish. 
As a result, the conditions \eqref{dilaton-stabilization} and \eqref{volume-stabilization} 
directly fix the dilaton and the volume fixed in terms of the six remaining independent $s^i$ moduli. 
Indeed we find 
\be
\label{dilaton-SOL}
e^{-\phi} = - {\cal G}^{-1} \int \star \Phi \wedge F_3 \,, 
\ee
and 
\be
\label{vol-AdS}
\text{vol}(X)^2 = - \left( \frac{6}{7 W_1} \right)^2 \, {\cal G} \, \int \star \Phi \wedge F_3 \,. 
\ee
Then the seven conditions \eqref{pi27-cond} fix the six remaining $s^i$. 
However, 
since they are readily contained in the first equation in \eqref{S-dual-susy}, 
we can work directly with the latter. 
We can recast in fact the first equation in \eqref{S-dual-susy} to take the form 
\be
\label{ADS-EQ-33}
\frac{6  {\cal G} \star F_3  }{7 W_1 \text{vol}(X)} + \frac76 W_1 \star \Phi =  d \Phi \,, 
\ee
which then explicitly gives seven equations once we expand on the $\Psi_i$ basis. 
After some manipulations these equations read 
\be
\label{F3-eq-AdS}
\frac{6 {\cal G}}{s^i} \left( \frac{f^i}{s^i} - \sum_{k=1}^7 \frac{f^k}{s^k} \right) 
= \frac{\left( \sum_{m,n} s^m {\cal M}_{mn} s^n \right)}{\left( \prod_l s^l\right)^{1/3}} \sum_j s^j {\cal M}_{ji} \, ,  
\ee
and should be solved in terms of the $s^i$. 
Clearly \eqref{F3-eq-AdS} (or equivalently \eqref{ADS-EQ-33}) 
describes only six independent equations due to the condition \eqref{vol-AdS}
that is satisfied by the volume. 
We conclude that the RR and geometric fluxes give the possibility to stabilize all 8 moduli 
on a supersymmetric AdS$_3$ vacuum.

An example for the matrix ${\cal M}_{ij}$ (this is a specific instance of the $SO(p,q) \times U(1)$ example of \cite{DallAgata:2005zlf}) 
that leads to full moduli stabilization is 
\begin{equation}
\label{M-Example-AdS-MASSIVE}
{\small
  {\cal M} = \begin{pmatrix}
    0 & 0 & 0 & h & 0 & 0 & m  \\
    0 & 0 & 0 & h & 0 & 0 & m \\
     0 & 0 & 0 & -h & 0 & 0 & m \\
    h & h & -h & 0 & 0 & 0 & 0  \\
     0 & 0 & 0 & 0 & 0 & 0 & 0 \\
     0 & 0 & 0 & 0 & 0 & 0 & 0 \\
   m & m & m & 0 & 0 & 0 & 0    \\
  \end{pmatrix} , 
  }
\end{equation} 
and we further assume that there exists a supersymmetric AdS vacuum at the positions 
\be
\label{ANSA-GEN}
\langle \tilde s^i \rangle = 1 \ , \quad \langle \phi \rangle = \phi_0 \ , \quad \langle v \rangle  = v_0 \,. 
\ee
To find this solution we start by evaluating $P_\phi=0$ on the Ansatz \eqref{ANSA-GEN} 
and find that it is solved by 
\be
\label{dil-GEN}
{\cal G} = - e^{\phi_0 - \frac{v_0}{\sqrt{7}}} \left( \sum_i f^i \right) \,. 
\ee
Similarly, we can evaluate $P_v=0$ and using \eqref{dil-GEN} and the Ansatz \eqref{ANSA-GEN} to find 
\be
\label{v-GEN}
3 \left( \sum_i f^i \right) e^{ \frac{\phi_0}{2} + \frac{v_0}{2 \sqrt{7}} } = h + 3 m \,. 
\ee
This determines the explicit values for $\phi_0$ and $v_0$.
Then from the expressions $\partial P / \partial \tilde s^i=0$, 
for $i=1, \dots, 7$, 
we get a series of equations which we simply satisfy by assigning the appropriate values to the $f^i$. 
Once we make use of the Ansatz and the conditions \eqref{dil-GEN} and \eqref{v-GEN} the solutions read 
\be
\begin{aligned}
f^1 & = f^2 = -\frac23 e^{-\frac{\phi_0}{2} - \frac{v_0}{2 \sqrt{7}}} \, h \, , 
\\
f^3 &= \frac43 e^{-\frac{\phi_0}{2} - \frac{v_0}{2 \sqrt{7}}} \, h \, , 
\\
f^4 &= -\frac13 e^{-\frac{\phi_0}{2} - \frac{v_0}{2 \sqrt{7}}} \, (2 h - 3 m) \, , 
\\
f^5 &= f^6 =  \frac13 e^{-\frac{\phi_0}{2} - \frac{v_0}{2 \sqrt{7}}} \, (h + 3 m) \, , 
\\
f^7 &= \frac13 e^{-\frac{\phi_0}{2} - \frac{v_0}{2 \sqrt{7}}} \, (h - 6 m) \, . 
\\
\end{aligned}
\ee
One can also evaluate the vacuum energy which is given by 
\be
\langle V \rangle = - \frac{1}{144} e^{\frac{4v_0}{\sqrt{7}}} (h + 3 m)^2 \,. 
\ee
In general the above Ansatz/solution has full moduli stabilization on SUSY-AdS because 
\be
\text{det}[m^2] \ne 0 \,. 
\ee
For example, 
for a specific setup we can have 
\be
\phi_0 = -3 \, , \ 
v_0 = - 33 \sqrt{7} \, , \ 
h = 1 \, , \ 
m = -1 \, , 
\ee
which gives large volume and weak string coupling, 
and we can easily verify numerically that 
\be
\langle V \rangle <0  \ , \quad \text{Eigenvalues}[m^2] > 0 \,, 
\ee
which guarantees a full moduli stabilization. 
Note however that if we take $h = 3$ instead of $h = 1$ 
then we obtain a Minkowski solution with $\text{det}[m^2]=0$.

\subsection{Indication for scale separation}

Let us now discuss the possibility of having scale separation in the supersymmetric AdS vacua. 
The scalar potential in our setup has the form 
\be
\label{VwithT}
V = {\rm F}_0 \, e^{-20\beta v + \phi} 
- {\rm R}_0 \, e^{-16\beta v} 
+ {\rm T}_0 \, e^{-18\beta v + \frac{\phi}{2}}  
+ {\rm G}_0 \, e^{-28\beta v - \phi} \,, 
\ee
where ${\rm F}_0 = |F_3|^2/16$, ${\rm G}_0 = {\cal G}^2/16$, ${\rm T}_0 = - \mu_{O5}/8$ and ${\rm R}_0 = R_7/8$. 
Once we minimize \eqref{VwithT} we find the vacuum values $v_0$ and $\phi_0$, 
and the vacuum energy is given by \eqref{Vcc}. 
To study the scale separation we follow closely \cite{Farakos:2020phe}, 
which means we ask that we can have flux values such that there is a limit where 
\be
\label{scalee-sepp}
\frac{L^2_{\text{KK}}}{L^2_{\Lambda}} = e^{16 \beta v} \, V_{\text{vac}} \to 0 \,. 
\ee 
Here $L_{\text{KK}}$ is the Kaluza--Klein scale that characterizes the internal space, 
and $L_{\Lambda}$ is the scale that characterizes the external 3D Anti-de Sitter space. 
Further details for \eqref{scalee-sepp} can be found in \cite{Farakos:2020phe}.

To this end we  consider a scaling limit where ${\rm G}_0 \sim N^a$ and ${ \rm F}_0 \sim N^{a+2 b}$ as $N \to \infty$ and we demand that each term in the potential has the same scaling behavior. 
Equating the scaling for the internal and external flux terms implies
\be
N^{a+ 2 b} e^{\phi - 20 \beta v} \sim N^a e^{-\phi - 28 \beta v} 
\implies N^{ - 2 b}e^{-8 \beta v}  \sim   e^{2\phi} \equiv N^{2p} 
\, ,
\ee
which leads to 
\be 
V_{\text{vac}} \sim N^{a + 6 p + 7b}  \ , \quad T_0 \sim N^{a + p + \frac{5}{2} b} \ , \quad  R_0 \sim N^{a+2p + 3b} \ , 
\quad e^{16 \beta v} V_{\text{vac}} \sim N^{a+2p + 3b} \,. 
\ee
The fact that ${\rm R_0}$ has the same scaling as $e^{16 \beta v} V_{\text{vac}}$, means that in order to achieve scale separation, we need to be able to take ${\rm R_0}$ small. We can also see that ${\rm T_0}^2 \sim {\rm R_0 F_0}$, consistent with the supersymmetric origin of the O5 term.

In principle we expect our scaling limit to correspond to some large value for the fluxes. However,
the tadpole condition $d F_3 = - \mu_{O5} J_4$ means that
the only possible consistent scaling we can have is 
\be
{\rm T}_0  \sim  {\rm F}_0 \sim {\rm R}_0 \sim N^0 \, , 
\ee
making scale separation impossible due to flux quantization.

Thus, to achieve scale separation we have to first cancel the tadpole in such a way that 
the fluxes are not restricted neither from the Bianchi nor from the torsions. 
This will allow them to take parametrically large or small values independently. 
As a result we include D5s such that for the tadpole of the form \eqref{WithD5s} we get 
\be
d F_3 = 0 = - \mu_{O5} \, J_4(O5) - \mu_{D5} \, J_4(D5) \,, 
\ee
where we recall that $\mu_{O5} > 0$ and $\mu_{D5} < 0$, 
but we keep the $F_3$ flux to non-vanishing values, 
that is $F_3 \ne 0$ and $F_7 \ne 0$. 
This can be achieved by taking 
\be
\label{fMzero}
\sum_i f^i d\Phi_i = 0 \,, 
\ee
which can have non-trivial solutions due to the freedom in choosing the torsion. 
Returning to the scalar potential, which is now missing the contribution 
from the O5-plane as it is cancelled by the D5-branes, we have 
\be
V = {\rm F}_0 \, e^{-20\beta v + \phi} 
- {\rm R}_0 \, e^{-16\beta v} 
+ {\rm G}_0 \, e^{-28\beta v - \phi} \,. 
\ee
Note also that the supersymmetric minimization with respect to $\phi$ and $v$ is bound to give \eqref{Vcc}, 
that is 
\be
\label{VAC-EX}
\langle V \rangle = - \frac{{\cal G}^2}{16} e^{-\phi_0} \text{vol}(X)^{-4} = - {\rm G}_0 e^{-28\beta v_0 - \phi_0}  \,. 
\ee  
This also guarantees that our moduli stabilization is consistent and non-tachyonic. 
Equivalently one can vary the scalar potential with respect to the volume and the dilaton to get 
\be
\label{vac-sol-ex}
 3 \, {\rm G}_0 \, e^{-28\beta v - \phi} = {\rm R}_0 \, e^{-16\beta v}  
 \ , \quad 
 {\rm F}_0 \, e^{-20\beta v + \phi}  = {\rm G}_0 \, e^{-28\beta v - \phi} \, , 
\ee
which are consistent for this setup and give again the supersymmetric vacuum energy \eqref{VAC-EX}. 
To obtain scale separation, we take the scaling\footnote{Small values for the torsion can be also used, 
as in \cite{Cribiori:2019hrb} and \cite{Andriot:2020wpp}, 
to get de Sitter vacua. 
For constraints on geometric fluxes see e.g. \cite{Marchesano:2006ns}.} 
\be
R \sim N^{-2}  \ , \quad F_3 \sim N^0 \ , \quad F_7 \sim N^0 \,. 
\ee
Asking that all the terms in the scalar potential scale in the same manner, we get 
\be
V \sim N^{-6} \ , \quad g_s = e^{\phi} \sim N^{-1} \ , \quad \text{vol}(X) = e^{7 \beta v} \sim N^{\frac74} \,. 
\ee
Then we finally get 
\be
\frac{L^2_{\text{KK}}}{L^2_{\Lambda}} = e^{16 \beta v} \, V \sim N^{-2} \,. 
\ee 
We conclude that achieving parametric scale separation requires taking the internal curvature $R_7$ to extremely small but positive values. Note that in this limit we remain at weak string coupling and large volume, 
and are therefore well within the regime of validity of the supergravity approximation. 
Since the $\tilde s^i$ should be fixed to finite values (otherwise the volume becomes singular), 
requiring small internal curvature means we have to tune the structure constants in the twisted torus. 
However, this requirement can run into tension with quantization conditions on the structure constants \cite{Marchesano:2006ns}, 
potentially making parametric scale separation impossible for the types of compactifications considered here. 
For further discussion on the intricacies of achieving scale separation in string theory see e.g. 
\cite{DeWolfe:2005uu,Polchinski:2009ch,Emelin:2020buq,Gautason:2015tig,Tsimpis:2012tu,Font:2019uva,Lust:2020npd,Buratti:2020kda,Petrini:2013ika,Alday:2019qrf,Marchesano:2020qvg,Andriot:2020vlg}. 
Note in particular that in \cite{Petrini:2013ika} the difficulty to get scale separation in IIB vacua has been anticipated.

\section{Brane supersymmetry breaking} \label{BSB}

\subsection{Introducing anti-D9s}

Until now we have worked with O5-planes which due to the orbifold involutions gave rise to an image O9-plane, 
that is an object with negative tension and with charge with opposite sign than that of a D9-brane. 
However, 
instead of a conventional O9 one can consider a so-called ``O9$^{+}$'' which has \emph{positive} tension, 
and charge with the same sign as that of a D9-brane. 
Then the RR tadpole for the O9$^{+}$ is now to be cancelled by 16 anti-D9-branes. 
This combination of $\text{O9}^{+}/\overline{\text{D9}}s$ is the so-called 
brane-supersymmetry-breaking (BSB) setup (a very recent review can be found in \cite{Mourad:2017rrl}). 
The gauge theory on such setup is USp(32), 
and because both the anti-D9-brane and the O9$^{+}$ have positive tensions, 
these add up and give a non-vanishing dilaton-dependent vacuum energy. 
In addition supersymmetry on the world-volume of this system is spontaneously broken and non-linearly realized. 
In particular the vacuum energy in the 10D Einstein frame has the form 
\be
V_{\text{10D-BSB}} = {\rm B}_0 \, e^{\frac{3}{2} \phi} \,. 
\ee 
In \cite{Dudas:2000nv,Pradisi:2001yv} for example the coefficient ${\rm B}_0$ is specified to be ${\rm B}_0 = 64 T_9$, 
where $T_9$ is the D9-brane tension up to the $e^{\frac{3}{2} \phi}$ dilaton factor, as in \eqref{DPOP}. 
Once we perform a direct dimensional reduction by inserting our metric Ansatz \eqref{10DmetricV} it becomes in 3D 
\be
V_{\text{3D-BSB}} = {\rm B}_0 \, e^{-14 \beta v +\frac{3}{2} \phi} \,. 
\ee
To embed this new term in the 3D superpotetnial we have to include a real scalar nilpotent superfield \cite{Buchbinder:2017qls}, 
let us call it $X$, 
which satisfies 
\be
X^2=0 \,. 
\ee
As in 4D (see e.g. \cite{Bergshoeff:2015jxa}) such nilpotent superfields tend to rise the vacuum energy and capture the 
effects of anti-branes. 
The modification to the $G_{IJ}$ metric to account for the coupling of $X$ to 3D supergravity will be $G_{XX} = 1$ 
and $G_{Xi} = G_{Xv} = G_{X\phi} = 0$, 
whereas the superpotential contribution is 
\be
P^{\text{BSB}} = \sqrt{{\rm B}_0} \, X \, e^{-7\beta v + \frac{3}{4}\phi} \,. 
\ee
Let us stress that this non-linearity is intrinsic and it 
is inherited directly by the non-linear supersymmetry of the 10D BSB theory  \cite{Antoniadis:1999xk,Angelantonj:1999ms,Dudas:2000nv,Pradisi:2001yv}.

With the inclusion of the BSB term the total scalar potential for the volume-dilaton sector, 
i.e. ignoring the $\tilde s^a$ or assuming they are stabilized, 
reads 
\be
\begin{aligned}
\label{V-stab-shape}
V &= {\rm F}_0 e^{ - 20\beta v + \phi } + {\rm G}_0 e^{ - 28 \beta v - \phi  } 
+ {\rm T}_0 e^{- 18\beta v + \frac{1}{2} \phi    }- { \rm R}_0 e^{- 16 \beta v} + {\rm B}_0 e^{- 14 \beta v + \frac{3}{2} \phi } \\
&\equiv  F + G + T - R + B \,. 
\end{aligned}
\ee
With ${\rm F_0,G_0,B_0} \geq 0$ and ${\rm T}_0 \leq 0 $ and we temporarily change notation such that $F = V^{F3}, G = V^{F7}$ etc. for visual convenience in the equations below.
A critical point of this potential satisfies
\be 
\begin{aligned}
\label{gradV}
 4F + 14 G + 5 T - 6 R =& 0 \, , 
 \\
 2F - 2 G + T + 3 B =& 0 \,. 
\end{aligned}
\ee
Which allows us to express the vacuum energy as
\be
\label{VBG}
V_{\text{vac}} = \frac{ B}{2} - G \,. 
\ee
The dependence of the vacuum energy only on two terms instead of three is remarkable. Solving \eqref{gradV} for different pairs of terms and substituting back into the potential results in an apparent dependence on all three remaining terms. However, note that the potential with only $F, T, R$ terms would result in a no-scale or runaway potential, and thus vanishing vacuum energy, while the other terms give additive corrections to the scalar potential, without generating cross-terms. In other words, the cosmological constant is ultimately determined solely by the interplay of Freund--Rubin-type fluxes ($F_7$) and SUSY breaking terms. 
The reason internal fluxes do not contribute to the cosmological constant appears to be that satisfying the tadpole condition by O5 planes generates precisely the right tension to cancel their contribution. This is in line with the observation that although O-planes appear to evade the usual supergravity de Sitter no-go theorems \cite{Gibbons:1984kp, Maldacena:2000mw}, once the flux they source is taken into account, the total stress-tensor does not produce a positive contribution to the vacuum energy \cite{Dasgupta:2014pma}.\footnote{From 
\eqref{VBG} we also see why one cannot get de Sitter vacua from Type I by simply adding anti-D5s, 
and instead we have to switch to the BSB setup to be able to even discuss such possibility. 
The fact that one may need two types of supersymmetry breaking sources to get classically stable de Sitter vacua was already 
alluded to in \cite{Farakos:2020idt} and as we will see we will need here both anti-D5s and anti-D9s as well, 
once we discuss the shape moduli stabilization.} 

While in the case of the Freund--Rubin term \eqref{PF7} the lack of additional cross-terms is justified by supersymmetry, 
with the BSB term, 
this amounts to ignoring backreaction from the anti-branes and therefore constitutes an important caveat to the analysis. 
It is possible that additional backreaction terms in the spirit of \cite{Moritz:2017xto} are present. 
Nonetheless, let us press forward and explore the possibility of de Sitter minima of this potential. 
The mass matrix eigenvalues are 
\be
m^2_{\pm} =  \frac{1}{7} \bigg( 20 G - T - 2 B \pm \sqrt{88 B^2 - 3B(8 G + T) + (8 G + T)^2   } \bigg) \, , 
\ee
which are positive when
\be
B < 8 G \ , \quad 
T < \frac{ 12 B^2 + 8 B G - 48 G^2  }{ B - 8G } \,. 
\ee
Note that this in principle allows for a positive vacuum energy when
\be \label{window}
2G<B<8G \implies 
2<\frac{ {\rm B_0} }{ {\rm G_0}} e^{ 14 \beta v + \frac{5}{2}\phi}<8 \, , 
\ee
with the lower inequality giving positive energy, while the upper inequality guaranteeing (meta-) stability. 
Note however that this imposes a relation between the stabilized values of the dilaton and the volume. 
Finally, 
our findings are consistent with \cite{Basile:2020mpt} because we have O5 sources. 
However the final verdict on the existence of such de Sitter vacuum can only be made after 
we stabilize the $\tilde s^a$ moduli and we take into account flux quantization.

As in the supersymmetric case, we can consider a scaling limit where ${\rm G}_0 \sim N^a$ and ${ \rm F}_0 \sim N^{a+2 b}$ as $N \to \infty$ and we demand that each term in the potential has the same scaling behavior. 
This once again determines the scalings 
\be
\quad V_{\text{vac}} \sim N^{a + 6 p + 7b} \ , 
 \qquad B_0 \sim N^{a + p + \frac{7}{2}b} \ , \qquad T_0 \sim N^{a + p + \frac{5}{2} b} \ , \qquad  R_0 \sim N^{a+2p + 3b} \, , 
 \ee 
 which lead to 
 \be
\frac{B_0}{G_0} e^{14 \beta v + \frac{5}{2} \phi} \sim N^{-b} \ , \qquad e^{16 \beta v} V_{\text{vac}} \sim N^{a+2p + 3b} \, , 
\ee
where we note that $b \neq 0$ means that we inevitably violate \eqref{window} as $N\to \infty$. This means that we need ${\rm F_0} \sim {\rm G_0} $ to preserve the stable de Sitter vacua. As before, ${\rm R_0}$ has the same scaling as the scale-separation parameter, $e^{16 \beta v} V_{\text{vac}}$, so we need to be able to take it small to achieve parametric scale separation, conflicting with the quantization of geometric flux. Furthermore, $b=0$ also ensures that ${\rm B_0}$ and ${ \rm T_0}$ have the same scaling, which we expect due to both terms arising from branes.

In fact we may further demand ${\rm T_0} \sim {\rm B_0} \sim N^0$, which requires $p = - a$ and ${ \rm R_0} \sim N^{p}$. This does indeed become small at large internal volume and weak coupling, yielding scale separation, but being in tension with quantization of the structure constants of the internal manifold.

On the other hand, if we don't demand parametric scale-separation, i.e. ${\rm R}_0 \sim N^0$ then we have ${\rm B_0} \sim { \rm T_0} \to \infty$ in our scaling limit. This, however is also unacceptable since the magnitude of ${\rm B}_0$ is fixed.\footnote{This situation is similar to \cite{Banlaki:2018ayh}, where weakly coupled, large volume 4d dS compactifications of massive type IIA appear to require large numbers of O6 planes.} Thus despite the scalar potential appearing to have de Sitter critical points, string theory does not seem to allow for parameter values such that these critical points appear at large internal volume and weak coupling, 
where this scalar potential is trustworthy.\footnote{A similar effect 
can be observed directly in gauged 4D N=2 supergravity \cite{Cribiori:2020use}.}

\subsection{Explicit examples of 3D de Sitter solutions?}

Actually, 
achieving full moduli stabilization including the $\tilde s^a$ is challenging, 
and we do not have a systematic way of tackling this question. 
However it is instructive to see first if we can generate the de Sitter vacua with the co-calibrated G2 geometry we have at hand 
following the methodology we also followed in \cite{Farakos:2020idt}. 
This does not give the most general de Sitter solution but it offers a simple way to obtain it. 
First we want to stabilize the $\tilde s^a$ in their ``autonomous'' supersymmetric positions, 
which means supersymmetric position of $\tilde s^a$ which do not require to fix the other moduli. 
From \eqref{pi27-cond} we see that we would need 
\be
\label{hatsa}
W_{27} = 0 \ , \quad \pi^{27}(F_3)=0 \,, 
\ee 
such that the dilaton VEV is kept free and is to be determined independently. 
In addition the equation $W_{27} = 0$ can be also solved independent of the volume modulus. 
Indeed taking into account that $s^i = e^{3 \beta v} \tilde s^i$, 
equation $W_{27} = 0$ takes the volume-independent form 
\be
\label{tonipie}
\sum_i \tilde s^i M_{ij} - \frac{1}{7}\Big( \sum_{m,n}\hat s^m M_{mn} \tilde s^n \Big) \frac{1}{\tilde s^j} = 0 
\ , \quad \prod_i \tilde s^i =1 \,. 
\ee
Then we notice that due to the structure of the scalar potential, 
even when it includes the BSB term, we have 
\be
\label{hats-crit}
\frac{\partial P}{\partial \tilde s^a} \Big{|}_{\eqref{tonipie}} = 0 
\quad \to \quad 
\frac{\partial V}{\partial \tilde s^a} \Big{|}_{\eqref{tonipie}} =0 \,. 
\ee
This is because of the properties \eqref{PRRR} and \eqref{PFFF}, 
but also $(G^{\phi\phi})_a=(G^{vv})_a=0$, 
and of course from \eqref{PF7} we automatically have $P^{F7}_a \equiv 0$. 
Therefore we can find vacua where the shape moduli are stabilized at their autonomous SUSY positions, 
and then we need only to stabilize the volume and the dilaton. 
This is exactly how the stabilization happens in \cite{Farakos:2020idt}. 
Now let us see if under the assumptions \eqref{hatsa} we can get de Sitter. 
We do not have to go into details, 
only check if the conditions we derived for de Sitter solution still hold. 
From \eqref{gradV} and \eqref{VBG} we see that 
\be
R = -F -G -5 V_{\text{vac}} \,, 
\ee
which means de Sitter critical points exist only for $\tilde R^{(7)} < 0$. 
In contrast to the latter, 
we see that \eqref{hatsa} (or \eqref{tonipie}) dictates $\tilde R^{(7)} \geq 0$. 
We conclude that there do not exist any de Sitter critical points that 
can be found with the method we followed in \cite{Farakos:2020idt}. 
As we said this does not exclude the possible existence of de Sitter, 
however it does leave much less room for it.

The fact that the shape moduli interfere with the construction of de Sitter solutions 
has been also discussed for example in \cite{Danielsson:2012et,Bena:2018fqc}. 
Indeed we believe that our example shows exactly how fixing the shape moduli into their 
``autonomous'' supersymmetric positions creates problems to finding de Sitter. 
In other words, 
if we had the shape moduli fixed in such autonomous supersymmetric positions and then we tried to uplift 
the vacuum to de Sitter 
we would force them to move out of these supersymmetric positions, 
and so the stabilization procedure would have to be worked out from scratch. 
We conclude that one should not ignore the stabilization of shape moduli during the uplift, 
nor take it for granted when searching for realistic examples.

One could try to construct de Sitter vacua with the 
shape moduli in their supersymmetric positions by including also anti-D5-branes. 
Let us see what would happen if we included such objects - 
assuming momentarily they can be included consistently in our setup. 
Their contribution to the Bianchi identity would be 
\be
\label{TAD-anti-D5s}
d F_3 = - \mu_{O5} \, J_4(O5) 
- \mu_{D5i} \, J_4(D5i) 
+ \mu_{\overline{D5}i} \, J_4(\overline{D5}i) \,, 
\ee 
where $\mu_{\overline{D5}i} < 0$, 
and the brane action (ignoring open string moduli) is 
\be
S_{\overline{D5}s} = \frac18 e^{- 18 \beta v + \frac{\phi}{2}} \int_{\text{3D}} \sqrt{-\tilde g_3} 
\sum_{\text{3-cycles}} 
\left[
\int_{\text{3-cycle}}  \tilde \Phi  
\int_{\text{4-cycle}}  \mu_{\overline{D5}i}  \, J_4(\overline{D5}i) \right] \, . 
\ee
Once we also take into account the O5s contributions to the 3D action, 
the net effect leads to the typical ``doubling'' of the anti-D5 terms due to 
\eqref{TAD-anti-D5s}. 
As a result, on top of all the previous contributions we had until now, 
we also have the additional term 
\be
\label{antiD5s}
2 \times V^{\overline{D5}} = 2 \times  \frac18 e^{- 18 \beta v + \frac{\phi}{2}} \sum_i \mu_i \tilde s^i 
\ , \quad \mu_i = - \mu_{\overline{D5}i} > 0 \,. 
\ee
Then the total scalar potential is 
\be
\label{V-DS-CONTR}
V = V^{\text{BSB}} + V^R + V^{F3} + V^{F7} + V^{O5/D5} + 2 V^{\overline{D5}} \,. 
\ee
Note that here $V^{O5/D5}$ refers to the same contribution we had in \eqref{TOT-tot}. 
If one wanted to assign to the smeared O5-plane its honest correct contribution it 
would be $V^{O5} - V^{D5} + V^{\overline{D5}}$, 
and this the reason for the ``doubling'' of $V^{\overline{D5}}$ in \eqref{V-DS-CONTR} 
as well as the cancellation of the D5 contribution. 
Assuming now that $\tilde s^7 = 1 / \prod_a \tilde s^a$, 
then we can have compatibility with an isotropic critical point of the shape moduli by requiring 
\be
\label{mu-mu}
\tilde s^a=1 \ , \quad \mu_i = \mu >0 \,. 
\ee
However here we would directly run into two problems if we wanted to get a de Sitter solution with our prescription from \cite{Farakos:2020idt}. 
First of all the term \eqref{antiD5s} evaluated on the \eqref{mu-mu} critical point 
clearly affects only the ${\rm T}_0$ term in the volume-dilaton scalar potential. 
Therefore it cannot change the fact that critical points still require $\tilde R^{(7)} < 0$ 
which as we said is not possible to achieve with the autonomous shape moduli stabilization. 
The second problem we would run into is that the tadpole \eqref{TAD-anti-D5s} 
in the presence of a background with $dF_3=0$ (which will probably be  forced on us by the requirements \eqref{hatsa}) 
will require various D5s for the cancellation of the O5 charge. 
Then clearly we cannot easily add anti-D5s as such system will be typically be inherently unstable.

As a means to escape the aforementioned issues we could still include anti-D5-branes 
but instead this time \emph{not} ask that the shape moduli to be stabilized in their autonomous supersymmetric positions. 
This gives some more freedom in the construction and allows to find de Sitter critical points, 
albeit possibly inconsistent once flux quantization is taken carefully into account. 
However, 
here we want to give a general overview/exposition of the possibilities rather than 
proving the existence of a {\it bona fide} stable de Sitter solution. 
We will therefore be more liberal with the flux quantization and brane/plane tension constraints, 
but will still require basic self-consistency. 
In particular we do not include D5s, such that there is no obvious instability, 
and we also want to satisfy the tadpole condition \eqref{TAD-anti-D5s}, without D5s.  
Taking into account that $dF_3=\sum_{ij} f^j {\cal M}_{ij} \Psi_i$, 
the tadpole takes the form 
\be
\label{TAD-DS-EX}
\sum_j f^j {\cal M}_{ij} + \mu_i = - \mu_{O5} < 0 \ , \quad \forall \, i \,. 
\ee 
This is forced on us by the fact that the O-planes have the same contribution to 
each cycle tadpole and therefore, since we do not have D5s, we need all tadpole contributions related 
to $dF_3$ and $\mu_i$ to take the same value - otherwise the existence of D5s is implied. 
We shall work with the geometric fluxes that give rise to a matrix of the form 
(this is a specific choice of 2-step nilpotent example of \cite{DallAgata:2005zlf}) 
\begin{equation}
\label{M-Example-dS}
{\small
  {\cal M} = \begin{pmatrix}
    0 & m & m & m & m & m & m  \\
    m & 0 & 0 & 0 & 0 & 0 & 0 \\
    m & 0 & 0 & 0 & 0 & 0 & 0 \\
    m & 0 & 0 & 0 & 0 & 0 & 0  \\
    m & 0 & 0 & 0 & 0 & 0 & 0 \\
    m & 0 & 0 & 0 & 0 & 0 & 0 \\
    m & 0 & 0 & 0 & 0 & 0 & 0    \\
  \end{pmatrix} , 
  }
\end{equation} 
and with $F_3$ flux of the form 
\be
f^i = a \left( 1, \omega, \omega, \omega, \omega, \omega, \omega \right)\,. 
\ee
The shape moduli are stabilized at the positions 
\be
\tilde s^i = \left( \omega^{-\frac67}, 
\omega^{\frac17}, \omega^{\frac17}, \omega^{\frac17}, \omega^{\frac17}, \omega^{\frac17}, \omega^{\frac17} 
\right)\,, 
\ee
whereas the volume and dilaton are stabilized at the positions $\phi_0$ and $v_0$. 
To get a de Sitter critical point we need to tune the geometric flux such that 
{\small
\be
m \!= - \frac{e^{\frac{\phi_0}{2} + \frac{v_0}{2 \sqrt 7} } 
\omega^{\frac{11}{7}} 
\left( 
a (5\!-\!30 \omega) 
+ 
\sqrt{
a^2 (33 + 4 \omega (177 \omega \!-\! 29)) \!
- \! 4 
\omega^{-\frac{12}{7}} {\cal G}^2 
(\omega\!-\! 1)(24 \omega \!+\! 1) 
e^{\frac{2 v_0}{\sqrt{7}} - 2 \phi_0}
}
\right)
}{2+48 \omega} .
\ee
} 
Note that our solutions will have five parameters that we can in principle choose independently, 
which are 
\be
a \, , \ 
\omega \, , \ 
{\cal G} \, , \ 
\phi_0 \, , \ 
v_0 \quad \text{(free parameters of the solution)}\, . 
\ee
For the anti-D5-brane tensions we now have 
\be
\mu_1 = - \frac{1}{10} \left( 
30 a m \omega 
+ 2 a^2 e^{\frac{\phi_0}{2} + \frac{v_0}{2 \sqrt 7} } \omega^{\frac{18}{7}} 
+ {\cal G}^2 \omega^{\frac{6}{7}} e^{\frac{5v_0}{2 \sqrt 7} -\frac{3\phi_0}{2}} 
+ 24 m^2 e^{-\frac{\phi_0}{2} - \frac{v_0}{2 \sqrt 7} } \omega^{-\frac47}  
\right) \, , 
\ee
and 
\be
\mu_{2,3,4,5,6,7} = - \frac{1}{10} \left( 
5 a m  
+ 2 a^2 e^{\frac{\phi_0}{2} + \frac{v_0}{2 \sqrt 7} } \omega^{\frac{11}{7}} 
- e^{-\frac{3\phi_0}{2} - \frac{v_0}{2 \sqrt 7} } \omega^{-\frac{11}7}  
\left( 
e^{\phi_0} m^2 
- {\cal G}^2 e^{\frac{3v_0}{\sqrt 7}} \omega^{\frac{10}{7}} 
\right)
\right) \, . 
\ee 
Note that $\mu_1$ is different that the rest, 
this is because they need to cancel the different contribution of the $dF_3$ flux in each tadpole, 
even though the O5 contribution is the same. 
Finally, 
we also tune the BSB contribution to take the form 
\be
{\rm B}_0 = \frac{1}{40} e^{\frac{v_0}{2 \sqrt 7} -\frac{5\phi_0}{2}} 
\omega^{-\frac{10}7} 
\left( 
6 e^{\phi_0} m^2 
+ 4 {\cal G}^2 \omega^{\frac{10}7}  e^{\frac{3v_0}{\sqrt 7}} 
- 7 a^2 \omega^{\frac{22}7} e^{\frac{v_0}{\sqrt 7} + 2\phi_0} 
\right) \,. 
\ee
Then using our Ansatz and the specific aforementioned values for the various coefficients one can check that 
\be
\frac{\partial V}{\partial \phi} = 0 \, , \quad 
\frac{\partial V}{\partial v} = 0 \, , \quad 
\frac{\partial V}{\partial \tilde s^a} = 0 \, , 
\ee
with the vacuum energy given by 
\be
V_{\rm vac} = \frac{1}{80} e^{\frac{4v_0}{\sqrt{7}}}  
\left( 
6 m^2 \omega^{-\frac{10}7} 
- {\cal G}^2 e^{\frac{3v_0}{\sqrt 7} - \phi_0} 
-7 a^2 \omega^{\frac{12}7} e^{\frac{v_0}{\sqrt 7} + \phi_0} 
\right)\,. 
\ee
Clearly the existence of de Sitter depends on the specific values one chooses. 
In addition one can check that if we ask that $\mu_i=0$ we are driven to an AdS vacuum, 
therefore the inclusion of the anti-D5s is crucial.

Let us now give a few \emph{numerical} examples. 
We can have 
\be
{\rm Example \,\, 1} \, : \ 
\phi_0 = -3 \, , \ 
v_0 = -3 \sqrt{7} \, , \ 
{\cal G} = 0.01 \, , \ 
a = 2 \, , \ 
\omega = 0.09985 \, , 
\ee
which give self-consistent values 
\be
\mu_i > 0 \ , \quad {\rm B}_0 > 0 \ , \quad \mu_{O5}>0 \,, 
\ee
and also satisfy \eqref{TAD-DS-EX}. 
Note that in this example $m\simeq-0.0029<0$. 
For this numerical example we also find 
\be
V_{\rm vac} \simeq  1.274 \times 10^{-13} 
\, , \quad 
\text{Eigenvalues}[V_{IJ}] > 0 \, , \quad I=\phi,v,\tilde s^b \,, 
\ee
implying a stable de Sitter critical point. 
Clearly from the values of the various coefficients we see that this example is in sharp 
contradiction with all sorts of flux quantization conditions 
but also clearly the values of $\mu_{O5}$, $\mu_{\overline{D5}}$ and ${\rm B}_0$ are unrealistic. 
In addition from the values of $\phi_0$ and $v_0$ we see that we are definitely not safely within the large volume regime, 
however the string coupling is indeed small.

Another numerical example is to have 
\be
{\rm Example \,\, 2} \, : \ 
\phi_0 = -3 \, , \ 
v_0 = -33 \sqrt{7} \, , \ 
{\cal G} = 0.01 \, , \ 
a = 10 \, , \ 
\omega = 0.0998 \, , 
\ee
which still gives self-consistent values for the various coefficients and allows for slightly more realistic values for $\mu_{O5}$, $\mu_{\overline{D5}}$ (but still overall unrealistic). 
We see that we are now safely within a weak coupling and large volume regime, but flux quantization is clearly not taken into account. 
For this example we find 
\be
V_{\rm vac} >0 
\, , \quad 
[V_{IJ}] < 0 
\, , \quad 
I=\phi,v,\tilde s^b \,, 
\ee
therefore there are tachyons in the scalar sector.

We conclude that it seems that one can achieve (stable) de Sitter critical points from an effective theory model building perspective,
but the required coefficients seem to be totally unrealistic from the string theory perspective. 
However, 
we believe that one needs to do an exhaustive scan over the various parameter values that are 
allowed by string theory in order 
to give a final verdict on the existence of classical 3D de Sitter vacua in string theory and on their stability. 
Our aim here was instead to highlight these open possibilities 
and we leave an exhaustive investigation for de Sitter solutions to future work. 
We expect that the study of 3D de Sitter vacua can further contribute to 
our understanding of such vacua from the perspective 
of the swampland program \cite{Farakos:2020idt,Danielsson:2018ztv,Obied:2018sgi,Andriot:2018wzk,Andriot:2019wrs}.

We finally stress that even if perturbative stability is achieved for the closed string moduli, 
including (anti) D5s can open up new decay channels, both perturbative and non-perturbative, in the open string sector, even if the various parameters are within a controlled string theory regime. 
Such instabilities may lead to very short lived vacua or completely destabilize them
(for a recent review and an extended discussion see e.g. \cite{Farakos:2020idt}).

\section{Outlook}

In this work we have studied flux compactifications of string theory down to three external dimensions 
and have highlighted properties that make them an interesting playground to test various swampland conjectures. 
Our primary motivation was to provide the tools for the construction  of the 3D N=1 supergravity, 
focusing in particular on the superpotential. 
Then we studied some simple examples that give us intuition for the vacuum structure. 
We focused in particular on discussing the possibility of having de Sitter and Anti-de Sitter vacua with scale separation and have seen how these vacua are allowed by the effective theory, but are hindered once we take into account proper quantization conditions as required in string theory.

As an outlook for future work we would like to discuss various possible extensions. 
One direction to expand on would involve a careful treatment of the open string sector, 
which we have mostly ignored here. 
This can be done in various ways. 
Firstly, as we have seen, 
it is un-avoidable to include O9 planes in this setup and so D9-branes also  have to be included. 
This means that one must study carefully the D9-brane sector which leads to a non-abelian gauge theory 
in 3 dimensions. 
In addition since  in principle  we would also need to include D5-branes 
these would further contribute to the non-abelian gauge sector 
on the 3D external space as well as give rise to extra scalar moduli. 
Overall one would need to include new contributions also to the superpotential to correctly describe these sectors. 
Note that this setup could offer the basis for constructing a 3D toy-model version of the 4D KKLT construction. 
Indeed, 
the non-abelian gauge theory may give rise to gaugino condensation in the 3D EFT and including anti-D5-branes 
can give rise to a putative ulplift mechanism similar to the KKLT model. 
This may be a worthwhile endeavor as it may help 
to  further understand the properties of de Sitter vacua in string theory, 
if such vacua truly exist, 
or simply a way to get more intuition about KKLT-type constructions. 
Along these lines one could also investigate the impact of Euclidean D-branes that wrap internal cycles, which we have ignored in the present work. 
These should give rise to non-perturbative contributions similar to the 4D case, however the absence of suitable non-renormalization theorems in 3D N=1 means their form is less constrained. On general grounds we can expect these contributions to take the form of non-perturbative exponentials dressed by a perturbative series in the moduli describing the volume of the wrapped cycle. It is also interesting to note that due to the dimensionality of the branes involved in our setup, it seems that the effects of gaugino condensation may differ in form from those of Euclidean D-branes, unlike the 4D scenario.

Another direction worth pursuing is to go beyond the co-calibrated toroidal G2 and include also the $W_{7}$ torsion. 
This is a very interesting development as it would allow to have more cycles in the theory 
and so more interesting backgrounds may be found. 
We have worked here only with toroidal orifolds, 
however, 
one does not essentially need to restrict oneself to this set of compactifications.
For example it would be important to study manifolds where the internal space allows warping, 
and this would also be important if one tries to build a 3D KKLT type of model, 
as we discussed earlier. 
Yet another direction to pursue would be finding the underlying un-smeared solutions of the 
orbifolds we discussed here. In a similar vein, one could also try to realize scenarios where the D5/O5 charge remains delocalized along the internal manifold but comes from topological flux and curvature terms in the D9/O9 worldvolume theory \cite{Dasgupta:1997cd}. These scenarios should be related to resolutions of the orientifold singularities and therefore have a richer topology, with the un-smeared orientifold solutions as a limit. Such a study would undoubtedly shed more light on the properties of O-planes and the consistency of working with the smeared solutions presented here.

Finally, 
one could try to classify all the 3D N=1 vacua that arise from flux compactifications on G2 with torsion 
and get important insight about the properties of the 3D swampland, 
especially by comparing to the dual 2D CFTs. 
Indeed, as we have seen (from the few sample examples we presented) the 3D N=1 low energy supergravity 
has a very rich vacuum structure, 
which however remains tractable due to its relatively simple ingredients. 
This means that a full classification of the \emph{classical} 3D vacua (de Sitter and Anti-de Sitter alike) 
can be done and a thorough investigation of their properties 
is possible, especially using more advanced methods as for example proposed in \cite{Comsa:2019rcz}.

\section*{Acknowledgements} 
We thank Keshav Dasgupta, Alex Kehagias, Luca Martucci, Augusto Sagnotti 
and Thomas Van Riet for very helpful discussions and correspondence. 
The work of ME and FF is supported by the STARS grant SUGRA-MAX.

\end{document}